\documentclass[12pt,letterpaper]{article}
\pdfoutput=1
\usepackage{soul}
\usepackage{jheppub}
\usepackage{amsfonts}
\usepackage{amsmath}
\allowdisplaybreaks[4]         
\usepackage{amssymb}   
\usepackage{euscript}       
\usepackage{color}      
\usepackage{subfig,tikz}
\usetikzlibrary{decorations.pathmorphing}
\usepackage{graphics}    
\usepackage[normalem]{ulem}
\usepackage{mathrsfs}

\newcommand{\bea}{\begin{eqnarray}}
\newcommand{\eea}{\end{eqnarray}}
\newcommand{\ba}{\begin{eqnarray}}
\usepackage{braket}
\newcommand{\ea}{\end{eqnarray}}
\newcommand{\nn}{\nonumber \\}

\newcommand{\beq}{\begin{equation}}
\newcommand{\eeq}{\end{equation} }
\newcommand{\beqa}{\begin{eqnarray}}

\newcommand{\eeqa}{\end{eqnarray}}
\newcommand{\beqar}{\begin{eqnarray*}}
\newcommand{\eeqar}{\end{eqnarray*}}

\newcommand{\be}{\begin{equation}}
\newcommand{\ee}{\end{equation}}

\newcommand{\rmax}{r_{\rm max}}

\newcommand{\td}{\text{d}}

\def\be{\begin{equation}}
\def\ee{\end{equation}}
\def\bea{\begin{eqnarray}}
\def\eea{\end{eqnarray}}

\def\nn{\nonumber}

\title{ \boldmath Chemistry and Complexity for Solitons in AdS$_5$}

\author{Shane Andrews,}
\author{Robie A. Hennigar,}
\author{and Hari K. Kunduri}

\emailAdd{saa837@mun.ca}\emailAdd{rhennigar@mun.ca}\emailAdd{hkkunduri@mun.ca }

\affiliation{Department of Mathematics and Statistics, Memorial University of Newfoundland.\\ St. John's, Newfoundland and Labrador, A1C 5S7, Canada \vspace{0.1cm}}

\date{\today}
\abstract{
Minimal $D=5$ supergravity admits asymptotically globally AdS$_5$ gravitational solitons (strictly stationary, geodesically complete spacetimes with positive mass).   We show that,  like asymptotically flat gravitational solitons, these solutions satisfy mass and mass variation formulas analogous to those satisfied by AdS black holes.  A thermodynamic volume associated to the non-trivial topology of the spacetime plays an important role in this construction. We then consider these solitons within the holographic ``complexity equals action'' and ``complexity equals volume'' conjectures as simple examples of spacetimes with nontrivial rotation and topology. We find distinct behaviours for the volume and action, with the counterterm for null boundaries playing a significant role in the latter case. For large solitons we find that both proposals yield a  complexity of formation proportional to a power of the thermodynamic volume, $V^{3/4}$. In fact, up to numerical prefactors, the result coincides with the analogous one for large black holes. 
 }

\begin{document} 
\maketitle
\flushbottom

\section{Introduction}
\label{sec:Introduction}
A classic result of four dimensional general relativity states that the only strictly stationary, asymptotically flat vacuum solution is Minkowski spacetime~\cite{Lich}. This result can now be obtained by an application of the positive mass theorem and Stokes' theorem along with identities satisfied by the stationary Killing field.  The result, which extends to the Einstein-Maxwell case, is often referred to as `no solitons without horizons' as black holes must be present in spacetime to achieve positive mass~\cite{Gibbons:1990um}.  If one  includes Yang-Mills fields, however, solitons do exist~\cite{Volkov:1998cc}, and allowing for different asymptotic behaviour also permits such solutions (e.g. $\mathbb{R}_t \times$ Euclidean Schwarzschild).  In higher dimensions, the existence of  asymptotically flat gravitational solitons has been investigated in great deal. Static solitons in Einstein-Maxwell theory can be ruled out~\cite{Kunduri:2017htl} and strictly stationary solitons can also be eliminated provided one assumes trivial topology~\cite{Shiromizu:2012hb}. However, many large families of solitons\footnote{More precisely, many of the supersymmetric solitons are not strictly stationary, but contain timelike hypersurfaces (`evanescent ergosurfaces')  along which the stationary Killing field becomes null. This property has been shown to to imply such solutions are nonlinearly unstable~\cite{Eperon:2016cdd, Keir:2016azt}.}  have been explicitly constructed in supergravity theories in $D > 4$ (see e.g. the review~\cite{Bena:2007kg}).  They are characterized by non-trivial spacetime topology (e.g. the spacetime contains cycles, or `bubbles'). In fact one may even explicitly construct spacetimes containing both solitons and black holes~\cite{Kunduri:2014iga}, which leads to continuous violations of black hole uniqueness even for supersymmetric, spherical black holes~\cite{Horowitz:2017fyg}.  Curiously, it can be shown that solitons and, more generally, black hole-soliton configurations, satisfy a Smarr-type relation and `first law of mechanics' mass variation formula~\cite{Kunduri:2013vka}.

Consider now asymptotically Anti de Sitter (AdS) gravitational solitons. It is known that AdS is the unique strictly stationary solution of Einstein's equations with a negative cosmological constant with spherical conformal boundary~\cite{Qing} (if one allows for a toroidal conformal boundary, the AdS soliton provides a non-trivial example~\cite{Galloway:2002ai}).  In sharp contrast,  if one includes a Maxwell field, examples of \emph{static}, asymptotically globally AdS solitons can be numerically constructed~\cite{Herdeiro:2016plq}.  This property seems closely linked to the fact that Laplace's equation in AdS allows admits regular solution for every multipole, unlike in Minkowski spacetime.  In minimal $D=5$ gauged supergravity, explicit examples of non-supersymmetric, asymptotically globally AdS$_5$ gravitational solitons were constructed by taking a limit of local black hole metrics~\cite{Ross:2005yj}.  These include a one-parameter family of cohomogeneity-one BPS solitons (see~\cite{Cassani:2015upa}{ for a thorough discussion of the regularity of the local BPS metrics first found in~\cite{Chong:2005hr}).  These solutions can be oxidized to ten dimensions to obtain local solutions to Type IIB supergravity. However, there are global obstructions to regularity which can only be remedied by imposing certain quantization conditions on the parameters.  In particular, while the BPS solutions mentioned above  are not regular as ten-dimensional solutions when lifted along $S^5$, the analysis of~\cite{Cassani:2015upa}  showed that they are regular when oxidized along more general Sasaki-Einstein spaces (e.g. $S^5 / \mathbb{Z}_3$).  In the present work we will mainly restrict attention to solitons as five-dimensional solutions of gauged supergravity. 

A natural question is whether such asymptotically AdS$_5$ solitons satisfy similar mass and mass variational formulas as do their asymptotically flat counterparts~\cite{Kunduri:2013vka}.  It is well known from the study of black holes that the standard method of deriving a Smarr relation (by applying Stokes' theorem to a Komar mass integral) does not work in the AdS setting because these integrals are divergent.  Over the last decade there has been a resurgence of interest in the thermodynamic properties of black holes in the presence of a (typically negative) cosmological constant. This interest was triggered by the observations of Kastor, Ray, and Traschen~\cite{Kastor:2010gq} that a consistent derivation of the Smarr formula for non-vanishing cosmological constant requires the introduction of a new quantity known as the thermodynamic volume, which is given by the integral of the Killing potential over the horizon.\footnote{Note that other authors had noticed aspects of this previously~\cite{Creighton:1995au, Caldarelli:1999xj, Kunduri:2005zg}, though the geometric origin and physical significance of the thermodynamic volume was not fully appreciated.}  The thermodynamic volume appears also in the first law of black hole mechanics as the conjugate potential to variations in the cosmological constant, interpreted in this context as a pressure. There has since been a relatively vast literature exploring the implications of the thermodynamic volume and pressure terms --- see~\cite{Kubiznak:2016qmn} for a review. While the majority of this work has focused on black holes,  it is worth emphasizing that the thermodynamic volume can arise from other sources as well. In~\cite{Mbarek:2016mep} a Smarr formula and associated first law was constructed for Eguchi-Hanson solitons in the presence of a positive cosmological constant. NUT charged spacetimes were considered in~\cite{Johnson:2014xza} where a first law and Smarr relation were constructed --- recently in~\cite{Bordo:2019tyh} it has been shown that in the Lorentzian version of these spacetimes the thermodynamic volume receives contributions from both the horizon and additionally from the integration of the Killing potential on the Misner strings. However, we note that in each of these cases there remains a horizon in the spacetime. 

In this article we consider a general class of cohomogeneity-one asymptotically globally AdS$_5$ gravitational solitons, invariant under a $\mathbb{R} \times SU(2) \times U(1)$ isometry group with four-dimensional orbits.  The spacetime is geodesically complete and contains a single 2-cycle (bubble) supported by magnetic flux~\cite{Ross:2005yj}.  We apply the above methodology to obtain a consistent mass formula for solitons. A careful application of Stokes' theorem reveals that the thermodynamic volume is non-trivial, despite the complete absence of horizons in the spacetime. Its geometric origin can be traced to a singular (but integrable) behaviour of the Killing potential at the location of the bubble. Considering the first law, the thermodynamic volume plays a role in the first law analogous to that in the black hole case. Interestingly, we find that the first law holds only if the variations respect the regularity of the spacetime, commensurate with previous results for the asymptotically flat case~\cite{Gunasekaran:2016nep}.

Our second application concerns holographic complexity. In recent years there has been considerable success derived from applying concepts in quantum information to problems in quantum gravity, in particular within the AdS/CFT correspondence. The quintessential example of this is entanglement entropy in CFTs which is deeply connected to the very existence of the bulk geometry~\cite{Ryu:2006bv, VanRaamsdonk:2010pw, Casini:2011kv, Lewkowycz:2013nqa,Hartman:2013qma}. Recently a new entry in the holographic dictionary has been proposed: circuit complexity. It has been argued that complexity will be necessary to capture certain features that entanglement cannot, such as the late-time behaviour of the black hole interior~\cite{Susskind:2014rva, Susskind:2014moa}. Intuitively speaking, complexity measures how ``difficult'' a state is to produce starting from a certain ``simple'' reference state and using only certain ``simple'' operations. Within the holographic context there have been a number of proposals for how the complexity of the CFT state is encoded in the bulk. The two most well-studied are the ``complexity equals volume'' and ``complexity equals action'' conjectures~\cite{Stanford:2014jda, Brown:2015bva, Brown:2015lvg}. The former relates the complexity of the CFT state to the volume of the extremal slice that meets the boundary at the timeslice on which the state is defined,
\be 
\mathcal{C}_\mathcal{V} = {\rm max} \left[\frac{\mathcal{V}(\mathcal{B})}{G \ell} \right] \, ,
\ee 
while the latter relates the complexity of the CFT state to the numerical value of the gravitational action evaluated on the Wheeler-DeWitt (WDW) patch, i.e. the domain of the dependence for bulk Cauchy surfaces that intersect the boundary at the time slice on which the state is defined,
\be 
\mathcal{C}_\mathcal{A} = \frac{I_{\rm WDW}}{\pi \hbar} \, .
\ee
There has been considerable effort dedicated to the exploration of these conjectures in the context of black holes. A number of appealing results have been obtained, including a simple, universal structure for the late-time growth of complexity of black holes in terms of thermodynamic quantities~\cite{Brown:2015lvg, Cai:2016xho, Huang:2016fks, Yang:2016awy, Cano:2018aqi, Jiang:2018pfk, Nally:2019rnw}. As of yet, there exists no first principles derivation for any of the holographic complexity conjectures, and moreover under a variety of circumstances their behaviour is qualitatively the same. It is therefore desirable to determine circumstances that robustly distinguish between the various conjectures as this may provide guidance in determining the proper behaviour.\footnote{Let us note that complexity in field theory is itself a relatively new area of exploration, especially in the regime of strong coupling. Nonetheless, proposals for circuit complexity in quantum field theory have yielded agreement with certain aspects of the action and volume conjectures, e.g. in the structure of UV divergences for free fields~\cite{Jefferson:2017sdb, Chapman:2017rqy}.} On the other hand, relatively little is known about the behaviour of these conjectures for spacetimes with rotation or nontrivial topology. In~\cite{Reynolds:2017jfs} the complexity conjectures were explored for the AdS soliton, finding both behaved monotonically as a function of the size of the soliton, but with the volume being strictly negative and the action positive. In~\cite{Bombini:2019vuk} the complexity equals volume conjecture was explored for microstate geometries finding monotonic behavior. It is our intention here to apply the holographic complexity conjectures to this class of AdS$_5$ solitons as simple examples of spacetimes with nontrivial rotation and topology. Some of authors have argued that the thermodynamic volume plays an important role in the understanding of complexity for black holes~\cite{Couch:2016exn, Fan:2018wnv, Liu:2019mxz,  Sun:2019yps}, and we will explore to what extent it does so here. Lastly, a number of recent studies have found that a certain counterterm for null boundaries is important to reproduce desired properties of the complexity in some circumstances~\cite{Chapman:2018dem, Chapman:2018lsv, Agon:2018zso, Alishahiha:2018lfv}, we will consider its implications for the soliton spacetimes.

Our paper is organized as follows. In section~\ref{globSol} we present the solution and discuss a number of its properties including its global structure, ergoregions, and geodesic motion. In section~\ref{solMech} we derive a mass formula for these solutions and present a first law of soliton mechanics. In section~\ref{holoComp} we apply the holographic complexity conjectures to the solitons. We conclude in section~\ref{discuss} with a discussion of the results. A number of appendices collect useful intermediate results.

\section{Solution and global structure}
\label{globSol}
\subsection{Metric, regularity \& conserved charges}

The bosonic part of the action of five-dimensional minimal supergravity is (setting Newton's constant $G_5 =1$ for the time being)
\begin{equation}\label{minsugra}
I = \frac{1}{16\pi} \int_\mathcal{M}  \star\left (R + \frac{12}{\ell^2}\right)  -2 F \wedge \star F -\frac{8}{3\sqrt{3}} F \wedge F \wedge A  \, .
\end{equation} Here $F = \td A$ and $A$ is a locally defined gauge potential. The existence of a non-trivial second homology $H_2$ implies that $F$ is closed but not exact.  The theory can be recovered from the general theory considered in \cite{Kunduri:2013vka} upon setting $I=1$, $g_{IJ}=2$ and $C_{IJK} = 16/\sqrt{3}$.  Minimal supergravity  can be obtained as a consistent truncation of Type IIB supergravity reduced on $S^5$.  The equations of motion following from this action are 
\begin{equation}\label{EFE}
\begin{aligned}
&R_{ab}. = 2 \left(F_{ac}F_{b}^{~c} - \frac{1}{6} |F|^2 g_{ab} \right) - \frac{4}{\ell^2}g_{ab} \, ,  \\
& \td \star F + \frac{2}{\sqrt{3}} F \wedge F =0 \, .
\end{aligned}
\end{equation}  The maximally symmetric solution of these equations is AdS$_5$ with radius of curvature $\ell$ and vanishing Maxwell field.  The equations of motion admit a 2-parameter family of smooth, stationary and biaxisymmetric  asymptotically AdS$_5$ horizonless solutions with positive mass, which we refer to as asymptotically AdS gravitational solitons.   

It is useful to first introduce some scalar potentials associated with stationary, biaxisymmetric ($\mathbb{R} \times U(1)^2$-invariant) solutions $(M,g)$ of the system \eqref{EFE}.  Denote by $\xi$ the stationary Killing vector field (normalized so that $|\xi|^2 \sim -r^2$ as $r \to \infty$ at the conformal boundary) and by $m_i$ the generators of the $U(1)^2$ action with $2\pi-$periodic orbits.  Closure of $F$ and simple connectedness of the spacetime implies the existence of globally defined potentials:
\begin{equation}\label{pot}
\td \Phi_\xi = i_{\xi} F, \qquad \td \Phi_i \equiv i_{m_i} F
\end{equation} where we always assume a gauge where $\Phi_\xi, \Phi_i \to 0$ at spatial infinity.  From the Maxwell equation we also obtain a closed 2-form
\begin{equation}
\Theta = 2 i_\xi \star F - \frac{8}{\sqrt{3}} F \Phi_\xi \, .
\end{equation} In the case of gravitational solitons, $\Theta$ need not be exact (since $H_2(M)$ is non-trivial) but we may construct globally defined potentials by noting that $L_{m_i} \Theta = \td (i_{m_i} \Theta) =0$ which guarantees the existence of potentials
\begin{equation}\label{Ui}
\td U_i = i_{m_i} \Theta = 2 i_{m_i} i_\xi \star F - \frac{8}{\sqrt{3}} \td \Phi_i \Phi_\xi.
\end{equation}  There are other potentials one may construct of this form although they will not be required here.  We may reduce stationary, biaxisymmetric solutions of \eqref{EFE} to a set of scalar functions defined on the orbit space $\mathcal{B} = M \setminus (\mathbb{R} \times U(1)^2)$. This is a two dimensional, simply connected space with an asymptotic end (corresponding to the asymptotic region of $M$) with a boundary corresponding to the fixed points sets of some integer linear combination of the $U(1)^2$ generators (these axes of symmetry are called `rods') and `corner' points where both generators vanish~\cite{Hollands:2007aj} (see also~\cite{Armas:2011ed}).  If a black hole is present then there will also be a rod corresponding to the event horizon where the $U(1)^2$ is free; we will not consider that case here.  It is natural to visualize $\mathbb{B}$ as the upper half plane $\{ (\rho,z)| \rho \in \mathbb{R}_+, z \in \mathbb{R}\}$ with $\rho =0$ corresponding to the boundary.  The topology of the spacetime is completely characterized by the rod data (e.g. a specification of the number of rods and a pair of integers associated to each rod indicating which combination of the $U(1)^2$ generators vanish upon it).

We will restrict attention in the present article to a 1 parameter family with an enhanced $\mathbb{R} \times U(1) \times SU(2)$ isometry group.  This soliton is obtained from the following three-parameter family of local solutions of \eqref{EFE}:
\begin{eqnarray}
\td s^2 &=& -\frac{r^2 W(r)}{4 b(r)^2} \td t^2 + \frac{\td r^2}{W(r)} + \frac{r^2}{4}(\sigma_1^2 + \sigma_2^2) + b(r)^2 (\sigma_3 + f(r) \td t)^2 \, , \\
F &=& \frac{\sqrt{3q}}{2} \td \left[\left(\frac{1}{r^2}\right)\left(\frac{j}{2}\sigma_3 - \td t\right)\right] \, ,
\end{eqnarray} where $\sigma_i$ are left-invariant one-forms on $SU(2)$:
\begin{align}
\sigma_1 &= -\sin\psi \td \theta + \cos \psi \sin \theta \td \phi \, , \quad \sigma_2 = \cos\psi \td \theta + \sin \psi \sin \theta \td \phi \,,
\nonumber\\
\sigma_3 &= \td \psi + \cos\theta \td \phi \, ,
\end{align}
which satisfy $\td \sigma_i = \tfrac{1}{2}\epsilon_{ijk}\sigma_j \wedge \sigma_k$.  The orientation is defined by the volume form $\td \text{Vol}(g) = \tfrac{r^3 \sin\theta}{8} \td t \wedge \td r \wedge \td \psi \wedge \td \theta \wedge \td \phi$.  For the solution to be asymptotically globally AdS$_5$ (the asymptotic region corresponds to $r \to \infty$) we require the identifications $ \psi \sim \psi + 4\pi$, $\phi \sim \phi +  2\pi$ and that $\theta \in (0,\pi)$.  The functions appearing in the metric are given by
\begin{eqnarray}
W(r) &=& 1 + \frac{4 b(r)^2}{\ell^2} -  \frac{2}{r^2}(p -q) + \frac{q^2 + 2 p j^2}{r^4} \quad f(r) = -\frac{j}{2 b(r)^2}\left(\frac{2p - q}{r^2} - \frac{q^2}{r^4}\right) \, ,\\
b(r)^2 &=& \frac{r^2}{4}\left(1 - \frac{j^2 q^2}{r^6} + \frac{2j^2 p }{r^4}\right) \, ,
\end{eqnarray} where $p,q,j \in \mathbb{R}$.   We will take $m_i = (\partial_{\hat{\psi}}, \partial_\phi)$, $\hat{\psi} = \psi/2$, to be our basis for the generators of the $U(1)^2$ action with $2\pi$-periodic orbits. 

The local solutions above can be made to describe rotating black holes by taking appropriate ranges of the parameters $(p,q,j)$.  Alternatively, as first observed by Ross~\cite{Ross:2005yj}, one may choose parameters so that a complete Lorentzian metric is obtained (the resulting solitons are AdS generalizations of the solitons considered in~\cite{Gunasekaran:2016nep}).  To do this, we require that the $S^1$ direction parameterized by the $\psi$ coordinate smoothly degenerates on an $S^2$ in the spacetime at some radius $r=r_0$.  Intuitively, spacetime near this $S^2$ is diffeomorphic to  $\mathbb{R} \times \mathbb{R}^2 \times S^2$ near the origin of the $\mathbb{R}^2$ parameterized by $(r,\psi)$~\cite{Gibbons:2013tqa}.  More precisely, we require that at some $r_0 > 0$, $b(r)^2$ and $W(r)$ have simple zeroes (this ensures $g_{tt} < 0$ at $r=r_0$).   The condition that they vanish requires
\begin{equation}
1 - \frac{2}{r_0^2}(p - q) + \frac{1}{r_0^4}(q^2 + 2pj^2) =0,  \qquad 1 - \frac{j^2 q^2}{r_0^6} + \frac{2 j^2 p }{r_0^4} =0, 
\end{equation} which implies that 
\begin{equation}
 p = \frac{r_0^4(r_0^2 - j^2)}{2 j^4} , \qquad q = -\frac{r_0^4}{j^2}. 
 \end{equation}  A necessary condition for the vector field $\partial_\psi$ to degenerate smoothly is to make the  identification 
 \begin{equation}
 \psi \sim \psi + \frac{4\pi}{\sqrt{W'(r_0)  (b^2(r_0))'}}
 \end{equation}  so that we must have $W'(r_0) (b^2(r_0))' = 1$ (one could choose $\psi$ to have period $4\pi/k, k \in \mathbb{Z}$, although then the spacetime would have a lens space $S^3/ \mathbb{Z}_k$ as its conformal boundary).  If we define a dimensionless parameter  $x \equiv r_0^2/j^2>0$, this condition becomes
 \begin{equation}\label{reg}
 (2+x)^2 ( 1 - x(1 - \alpha)) = 1 \, ,
 \end{equation} where $\alpha \equiv j^2/\ell^2>0$.  Hence $x(1 - \alpha) <1$ for a solution to exist.  This is easily seen to be equivalent to the requirement that $r= r_0$ is a timelike surface, namely 
 \begin{equation}
 \lim_{r \to r_0} g_{tt} = -(1 - x(1-\alpha)) < 0 \, .
 \end{equation} We write \eqref{reg} in the form
 \begin{equation}
 (1- \alpha) x^3 + (3 - 4\alpha) x^2 - 4\alpha x - 3 =0.
 \end{equation}  According to Descartes' rule of signs, the cubic will have exactly one positive root provided either $1 -\alpha > 0, 3 - 4\alpha > 0$ or $ 1 - \alpha >  0, 3 - 4\alpha <0$. One of these possibilities is achieved for $0 < \alpha < 1$, i.e. $0 < j^2 < \ell^2$. In the regime $j \to 0$ the solution approaches global AdS$_5$ (see appendix~\ref{smallSol}), while $r_0 \to \infty$ as $j\to \ell$.   For $\alpha =1$, there are no positive roots.  We note in passing that the Maxwell field is indeed regular near the smooth origin where $\partial_\psi$ degenerates (indeed it behaves like the volume form on $\mathbb{R}^2$).  Furthermore, $g(\partial_\psi, \sim)$ must vanish at the degeneration point in order to avoid Dirac-Misner strings (or else one must periodically identify $t$ and introduce closed timelike curves).  One can verify $b(r)^2 f(r) = O(r-r_0)$  and $b(r)^2 f(r)^2 = O(r-r_0)$ as $r \to r_0$ --- this is independent of the regularity condition \eqref{reg}.  Therefore the spacetime metric is smooth near the degeneration surface.

Before moving on, let us make a few further remarks on the structure of this spacetime.  
The determinant of the metric on a constant-$r$ surface is
\be 
g = -\frac{r^6 W(r)\sin^2\theta}{64} \, ,
\ee 
and therefore the absence of event horizons requires that $W(r) > 0$ for all $r > r_0$. This can be shown to be the case via the following argument.  First, note that $W(r)$ can be written as
\be 
W(r) = \frac{(r^2-r_0^2)}{\ell^2 r^4} \left[\ell^4 \alpha x (1 - x(1 + x(1-\alpha))) + \ell^2 (1 + \alpha x) r^2 + r^4 \right] \, .
\ee
To show that $W(r) > 0$ for all $r > r_0$ we must show that the term in square brackets is positive. For large enough $r$, it is clear that this term will be positive, and therefore positivity will hold in general provided that it has no zeros for $r > r_0$. Next write $r = r_0 \sqrt{(R + 1)}$ so that this term becomes
\be 
\ell^4 \left[x(2+x)(1-x(1-\alpha))\alpha + x \alpha (1 + 3 x \alpha) R + x^2 \alpha^2 R^2 \right] \, .
\ee 
The result $W(r) > 0$ for all $r > r_0$ will hold if this polynomial has no zeros for positive $R$. This follows from the rule of signs: each coefficient is positive, the first due to the regularity condition~\eqref{reg}, and the remaining two are manifestly positive.

The spacetime is also easily checked to be stably causal, i.e.~$g^{-1} (\td t, \td t) = -4 b^2/(r^2 W) < 0$ everywhere. This result follows from the previous considerations of $W$ along with the fact that $b^2$ is manifestly positive for $r > r_0$. Thus there are no event horizons and $t$ provides a global time function for the spacetime. We have therefore shown that provided $0 < j^2 < \ell^2$  one can find an $x_*>0$ so that $r_0^2 = x_* j^2$ will satisfy the regularity condition \eqref{reg}. In this case the local metric extends smoothly to a one-parameter family of  complete metrics on the underlying manifold $M =\mathbb{R} \times  \mathbb{CP}^2 \# \mathbb{R}^4$ (the complex projective plane with a point removed corresponding to the asymptotic end).  

The spacetime contains a non-trivial 2 cycle, or `bubble' located at $r=r_0$ which has topology $S^2$ and is equipped with its standard round metric of radius $r_0/2$.  This $S^2$ is supported by the dipole flux
 \begin{equation}
 q[C] \equiv \frac{1}{4\pi} \int_{S^2} F = \frac{\sqrt{3} r_0^2}{4 j}.
 \end{equation} We also record here the various scalar potentials associated to the Maxwell field:
 \begin{equation}\label{Max:Phi}
\Phi_\xi = \frac{\sqrt{3}q}{2r^2}, \qquad \Phi_{\hat\psi} = -\frac{\sqrt{3} q j }{2r^2}, \qquad \Phi_\phi = - \frac{\sqrt{3} q j \cos\theta}{4r^2}. 
\end{equation} A longer computation gives
\begin{equation}\label{Max:U}
U_{\hat\psi} = \frac{\sqrt{3} q j (q - r^2)}{r^4} , \qquad U_\phi = \frac{\sqrt{3} q j (q - r^2 - \frac{r^4}{\ell^2})\cos\theta}{2 r^4}. 
\end{equation} We have chosen integration constants so that the potentials $\Phi_\mu \to 0$ as $r \to \infty$ as well as $U_{\hat \psi}$. However, note that we cannot arrange for $U_\phi$ to vanish asymptotically due to the presence of the term involving the AdS length.   Indeed $U_\phi  \to -\sqrt{3} q j \cos\theta/ (2 \ell^2)$ as $r \to \infty$. This observation will play an important role in our construction of the mass formula for the spacetime.

The mass of these solutions can be obtained using the Ashtekar-Magnon-Das procedure which yields
\be 
M = \frac{\pi}{4} \left(3(p-q) + \frac{j^2 p}{\ell^2}\right) \, .
\ee
The same result is obtained using the holographic stress tensor --- see appendix~\ref{FGstress}. The soliton solutions considered here also carry non vanishing electric charge and angular momentum. The electric charge is defined by
\begin{equation}
Q \equiv \frac{1}{4\pi} \int_{S^3_\infty} \star F = \frac{\sqrt{3} \pi q}{2} = -\frac{\sqrt{3}}{2}\pi \left(\frac{r_0^4}{j^2}\right).
\end{equation}  We define the angular momenta with respect to the basis $m_i$ with $2\pi-$ periodic orbits via standard Komar integrals, which gives
\begin{equation}\label{angMom}
J[\partial_{\hat \psi}]  = \frac{\pi j}{4} (2 p -q) ,   \qquad J_\phi =0.
\end{equation} In terms of the angular momenta associated with generators that are orthogonal at infinity, these geometries have $J_1 = \pm J_2$ where the sign depends on the choice of orientation.   Supersymmetric solutions of minimal gauged supergravity must satisfy one of the following BPS conditions~\cite{Cvetic:2005zi}:
\begin{equation}\label{BPS}
M = \frac{\sqrt{3}}{2} Q + \frac{J_1}{\ell} + \frac{J_2}{\ell} , \qquad M = -\frac{\sqrt{3}}{2}Q \, .
\end{equation} Note that due to Chern-Simons term it is not possible to always arrange $Q \geq 0$ while simultaneously keeping $J_1 > 0, J_2 > 0$. Although the Einstein equation obviously is invariant under $F \to -F$y to preserve the Maxwell equation under this change,  we must also change the orientation.  For our class of solutions, we can show that there are no regular BPS solutions satisfying the first bound. This is consistent with the global analysis of a class of BPS solutions carried out in~\cite{Cassani:2015upa}.  Those authors considered the most general known BPS solutions and showed there were no 1/4 BPS solutions with $J_1 = J_2$.   However there  \emph{is} a 1/2 BPS solution with $SU(2) \times U(1)$ isometry satisfying the second bound.  This can be seen in our parameterization quite easily.  If we impose the second condition in \eqref{BPS} then we find a regular solution is given by
\begin{equation}
\alpha = \frac{1}{9}, \qquad x =1.
\end{equation} In terms of the original parameters, the BPS soliton corresponds to $r_0 = j = \ell/3$ where we have chosen $j  > 0$ without loss of generality. The resulting solution is the same as that given in \cite{Cassani:2015upa} although one has to flip the sign of both their Maxwell field and choice of orientation.  

\subsection{Ergoregions, geodesics, and trapping}
\label{ergoGeo}

Next we will discuss properties of the spacetime relevant to the motion of test particles and causal structure. Specifically, we will examine the parameter space for instances of ergosurfaces,  and discuss aspects of geodesic motion in the spacetime. Our discussion of ergoregions is consistent with that of~\cite{Compere:2009iy}, but is somewhat more detailed.

Let us begin by examining the solitons for the presence of ergoregions. These correspond to regions where the Killing field $\partial/\partial t$ becomes spacelike, which in turn requires
\be 
g_{tt} = -\frac{r^2 W}{4 b^2} + b^2 f^2 > 0 \, .
\ee
Regularity enforces that $g_{tt} < 0$ at the bubble, while asymptotically the metric approaches that of AdS$_5$ and so $g_{tt} < 0$ there as well.    Yet \textit{a priori} it is not obvious that the spacetime should be free from ergoregions, as seemingly nothing prevents the second term, which is manifestly positive away from $r = r_0$, from overtaking the first in some intermediate region.

To examine this more carefully, let us search for zeroes of $-g_{tt}$. Making the substitution $r = r_0 \sqrt{R + 1}$ along with $r_0^2 = x j^2$ and $j^2 = \alpha \ell^2$, we find that 
\be\label{ergPoly} 
-(1+R)^2 g_{tt} = 1 - x(1-\alpha) + \left(3 x \alpha + (1-x)(x+2) \right)R + (1+3 x \alpha) R^2 + x \alpha R^3 \, .
\ee
The regularity condition ensures that the first term is always positive, while the coefficients of $R^2$ and $R^3$ are clearly positive. The coefficient of the linear term is more complicated. The regularity condition only ensures its positivity provided
\be 
x_0 < x < x_1 
\ee
where $x_0 \approx 0.879385$ is the real, positive root of
\be 
-3 + 3 x_0^2 + x_0^3 = 0\, ,
\ee
and $x_1 \approx 1.49551$ is the largest real, positive root of
\be 
1 - 4 x_1 - 3 x_1^2 + 2 x_1^3 + x_1^4 = 0 \, .
\ee
The lower limit at $x = x_0$ corresponds to the limiting case where the soliton vanishes $r_0 \to 0$ --- regularity of the solution in this limit requires that $r_0^2/j^2 \to x_0$. Clearly, we cannot have $x < x_0$. However, the upper limit, $x = x_1$, does not correspond to any particular limit of the solutions as far as regularity is concerned. Indeed, the condition~\eqref{reg} permits $x$ to grow arbitrarily large, a result of the fact that $r_0 \to \infty$ as $j \to \ell$.  Therefore, provided that $x \ge x_1$ the cubic will have two sign flips. From the rule of signs we recognize that this could correspond to two or zero roots of the polynomial.

To see which situation is realized, we can examine the discriminant of the cubic polynomial,  treated as a function of $R$. The result is
\be 
\Delta = x^2 \left(4 \alpha  x^5-3 \alpha  (9 \alpha -4) x^4-6
   \alpha  x^3+(1-14 \alpha ) x^2+2 x-3\right)\, .
\ee
We can search for a solution of $\Delta = 0$ under the constraints that $x > x_1$ and subject to the regularity condition. We find that such a zero occurs when $x$ solves the following polynomial equation:
\be 
-48 + 104 x_\star + 5 x_\star^2 -142 x_\star^3 + 53 x_\star^4 + 38 x_\star^5 - 70 x_\star^6 - 20 x_\star^7 + 13 x_\star^8 + 4 x_\star^9 = 0\, ,
\ee
while the corresponding value of $\alpha$ is given  simply by
\be 
\alpha_\star = \frac{-3 + 3 x_\star^2 + x_\star^3}{x_\star(4 + 4 x_\star + x_\star^2) } \, .
\ee
Since the first equation is a ninth-order polynomial we cannot find a solution in closed form. Nonetheless, there is a single solution consistent with the various regularity conditions and it is easy to numerically approximate this solution, finding $x_\star \approx 2.10711$ with the corresponding $\alpha_\star \approx 0.553551$. Recall that the discriminant of a cubic vanishes at the point where the cubic has two, real coincident roots. We find that for $x > x_\star$ the discriminant is positive, indicating three distinct real roots. We know that two of these roots will always occur for $R > 0$ by the following reasoning. By applying the rule of signs to \eqref{ergPoly} with $R \to -R$, we conclude that for $x > x_\star > x_1$ there is a single sign flip, indicating the possibility of only a single zero for $R < 0$. Thus, at least one root must occur for $R > 0$. Moreover, we know from the asymptotics of $g_{tt}$ as $r\to r_0$ and $r \to \infty$ that any (positive) roots of $g_{tt}$ must come in pairs. Therefore, for $x > x_\star$ two roots must be located at positive $R$, i.e. for $r > r_0$.

\begin{figure}[htp]
\centering
\includegraphics[width=0.65\textwidth]{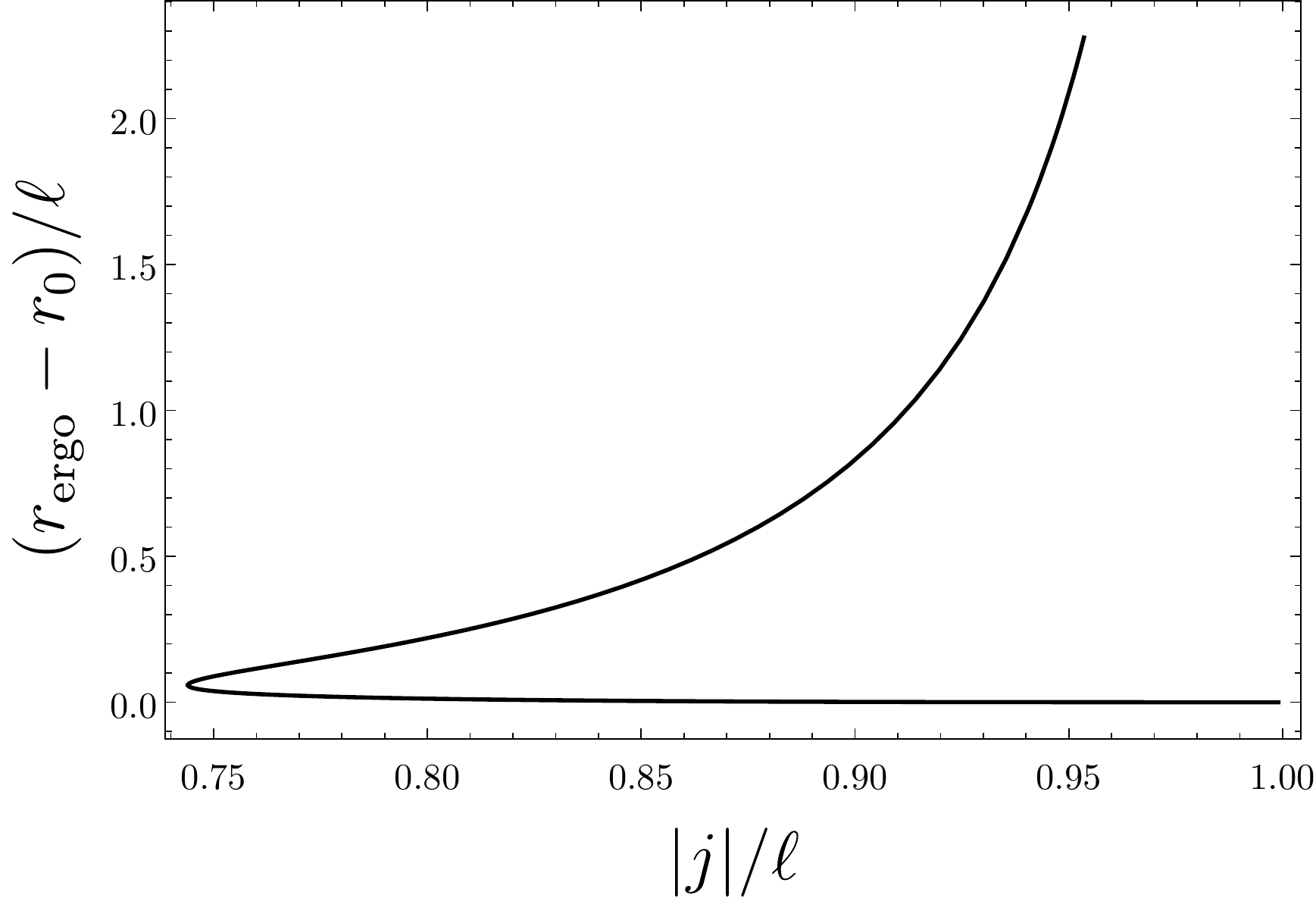}
\caption{A plot depicting the difference between the location of an ergosurface, $r_{\rm ergo}$, and the location of the bubble, $r_0$. For sufficiently large $j^2/\ell^2$, there are two distinct ergosurfaces. The inner ergosurface asymptotes to the location of the bubble from above as $|j| \to \ell$. }
\label{ergoRegions}
\end{figure}

The physical implications of these considerations are as follows. In the regime $x < x_\star$, $g_{tt}$ is strictly negative and no ergoregion is present. In particular, this reproduces the result that ergoregions are absent in the asymptotically flat versions of these solutions. For $x = x_\star$ $g_{tt}$ has a coincident zero. At this point $\partial/\partial t$ becomes null at a single values of $r$, never flipping sign. This corresponds to an \textit{evanescent ergosurface}. For $x > x_\star$, $g_{tt}$ has two zeroes for $r > r_0$. The vector $\partial/\partial t$ is spacelike between the zeros and timelike everywhere else, indicating two disjoint ergosurfaces with an ergoregion between. Note that regularity enforces that the $r = r_0$ surface is always timelike. Thus, all ergosurfaces must occur for $r > r_0$. Translating the conditions on $x$ and $\alpha$ back into these parameters we see that $x = x_\star$ corresponds to
\be 
\frac{r_0}{\ell} \approx 1.079995899 \, , \quad \frac{j}{\ell} \approx 0.7440098039 \, , \quad \frac{r_{\rm ev-ergo}}{\ell} \approx 1.138668419 \, .
\ee 
We see then that the evanescent ergosurface occurs for $r > r_0$. In general, the inner ergosurface is sandwiched between the evanescent ergosurface and the bubble, while the outer ergosurface occurs for a value of $r$ always larger than that corresponding to the evanescent ergosurface. This is illustrated in figure~\ref{ergoRegions}.  It is worth mentioning that the evanescent ergosurface is absent in the BPS solution (which corresponds to $r_0 = j = \ell/3$).  

These results suggest that for sufficiently large $j/\ell$, the AdS solitons considered here are unstable as they contain ergoregions.  Linear modes can have arbitrarily large negative energy, and become unbounded in time. In the special case where only an evanescent ergosurface exists, a nonlinear instability is expected to arise, due to a `clumping' of energy that provides an obstruction to sufficiently rapid decay of solutions to the linear wave equation. This in turn provides a mechanism for instabilities at the nonlinear level. This instability has been rigorously established in the asymptotically flat setting by Keir~\cite{Keir:2016azt, Keir:2018hnv} (see also the clear explanation given in \cite{Eperon:2016cdd}) and it seems plausible that a similar mechanism may prevail in asymptotically globally AdS spacetimes.  

Let us now briefly consider the motion of test particles on the soliton spacetime. We will be interested in null geodesics. These will feature in the discuss of the causal structure, but are also relevant in determining whether or not the solution exhibits the property of \emph{stable trapping} whereby null geodesics are confined to a finite region of space (unlike the well known $r = 3M$ photon sphere in the Schwarzschild geometry, these orbits are stable).  Spacetimes exhibiting stable trapping (without any evanescent ergosurfaces) are also  likely to be unstable, as the trapping of energy into finite regions tends to cause slow rates of decay of linear solutions to the massless wave equation.  This phenomenon has already been exhibited in the case of Kerr-AdS$_4$ black holes~\cite{Holzegel:2011uu}. We will show below that stable trapping occurs quite generally for the family of solitons considered here, even in the BPS case.   We leave a thorough analysis of the stability of these solutions to future work. 


Consider the trajectories of null geodesics $x^a(\lambda)$ where $\lambda$ is an affine parameter.  The equations for null geodesics are Liouville integrable (e.g. the configuration space is five-dimensional and there are 5 Poisson commuting functions associated to the Hamiltonian itself as well as conserved quantities associated to the $\mathbb{R} \times SU(2) \times U(1)$ isometry).   Denote differentiation with respect to $\lambda$ by an overdot.  The equations for a geodesic $x^a(\lambda)$ are easily obtained via the Hamilton-Jacobi method with the results
\begin{equation}
\begin{aligned}
\dot t & = \frac{8 b(r)^2}{r^2 W(r)} ( E + f(r) p_\psi), \qquad \dot \psi = -f(r) \dot t + \frac{2}{b(r)^2} p_\psi + \frac{8 \cot\theta}{r^2}(\cot\theta p_\psi - \csc\theta p_\phi),  \\
\dot \phi & = -\frac{8 \csc\theta}{r^2} (\cot\theta p_\psi - \csc\theta p_\phi), 
\end{aligned} 
\end{equation}  and 
\begin{equation}
\begin{aligned}
\dot r^2 &= V(r) \equiv \frac{16 b(r)^2}{r^2} (E + f(r) p_\psi)^2 - \frac{4 W(r) p_\psi^2}{b(r)^2}  - \frac{16 C W(r)}{r^2}  \, ,\\
\dot \theta^2 & = \frac{64}{r^4} \left[ C - (\cot\theta p_\psi - \csc \theta p_\phi)^2 \right] \, ,
\end{aligned}
\end{equation} where $C, p_\psi, p_\phi, E$ are constants of the motion.   

\begin{figure}[htp]
\centering
\includegraphics[width=0.65\textwidth]{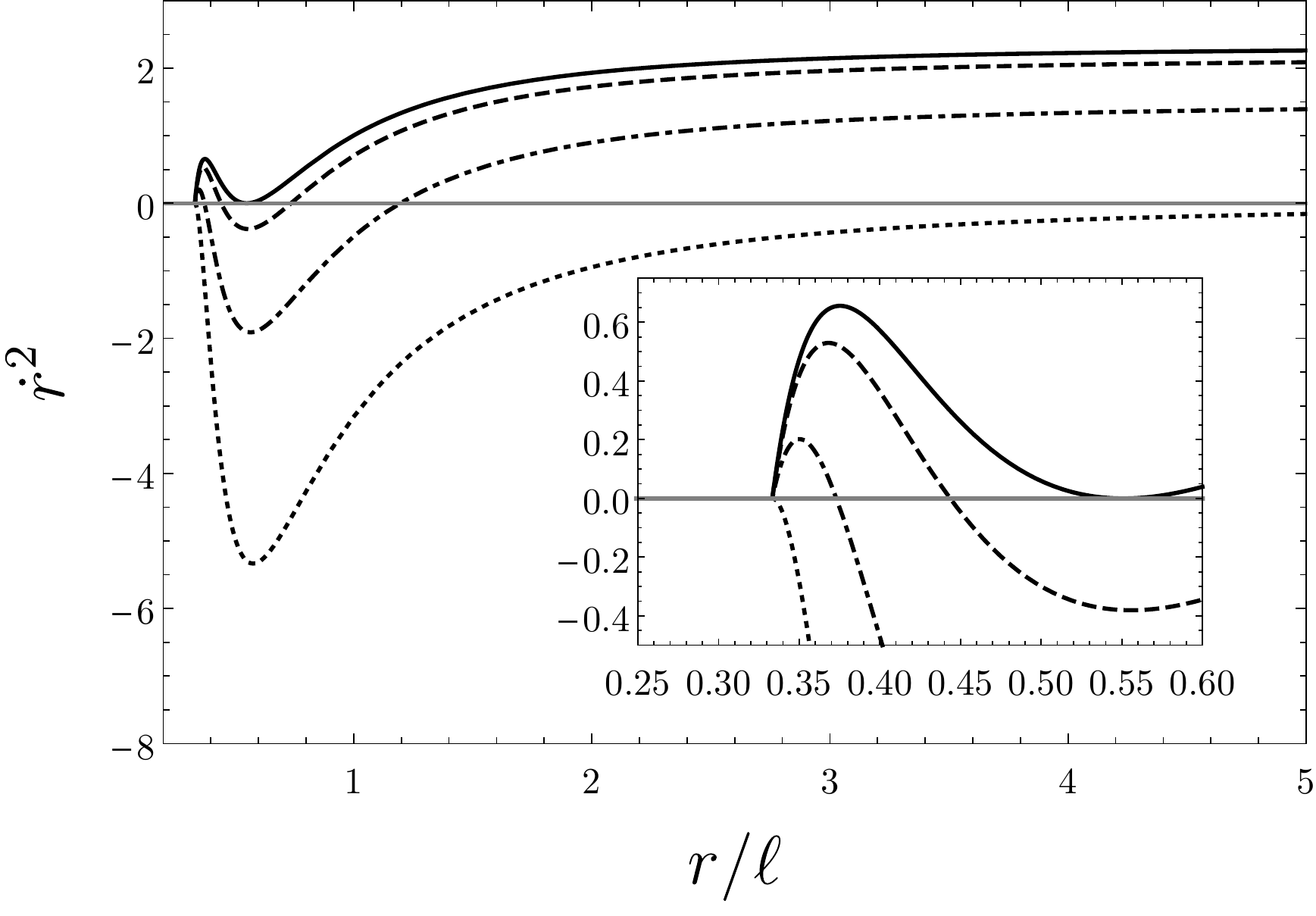}
\caption{Here we show the ``potential'' $V(r)$ as a function of radius for the BPS case $r_0 = j = \ell/3$. Turning points for geodesic motion correspond to zeros of $V(r)$. The curves (in order from top to bottom) correspond to $\mathscr{C} = \mathscr{C}_\star, 1.1 \mathscr{C}_\star, 1.2 \mathscr{C}_\star, 1/4$, where $\mathscr{C}_\star = (4 + 3 \sqrt{3})/88$. The first and last of these curves bound the region of parameter space for which trapping can occur. The inset zooms in on the the region where trapping can be seen.
}
\label{trapping}
\end{figure}

For simplicity, consider null geodesics with $p_\psi = 0$ and  lying on the equatorial plane with fixed $\theta(\lambda)  = \pi/2$.  This can be achieved by setting $C = p_\phi^2$.  Rescaling the affine parameter appropriately so as to absorb $E$, the radial equation reduces to
\begin{equation}
\dot{r}^2 = \frac{16}{r^2}\left[ b(r)^2  -  \mathscr{C} \ell^2 W(r)\right] \, ,
\end{equation}
 where $C =  \mathscr{C} \ell^2 E^2$ and $\mathscr{C}$ is a dimensionless constant parametrizing the angular momenta along the $\phi$ direction.  Stable trapping occurs when there is a region $[r_1, r_2]$ in which $\dot{r}^2 > 0$ in the interior and vanishes at the end points, with $\dot{r}^2 < 0$ immediately outside the closed interval.  Physically this means null geodesics must `turn back' at the points $r_1, r_2$.  It is fairly simple to see that stable trapping occurs generically even in this special region of of the configuration space of null geodesics. In particular, in the BPS case one can easily observe that provided $  \mathscr{C}_\star \equiv(4 + 3 \sqrt{3})/88  < \mathscr{C} < 1/4$  stable trapping will occur, as we illustrate graphically in figure~\ref{trapping}.  A full analysis of geodesics in this spacetime is beyond the scope of the present work, but it would be interesting to explore whether this trapping leads to a bound on the rate of decay of solutions to the linear wave equation. 


Finally, consider the case of `radial' null geodesics moving in the $(t, r, \psi)$ surface which have $C = p_\psi = p_\phi =0$.  For such geodesics we can fix $\theta = \pi/2$ and $\dot \phi =0$.  We can scale the affine parameter to fix $E=1$ so we have
\begin{equation}
\dot t = \frac{8 b(r)^2}{r^2 W(r)} , \qquad \dot r^2 = \frac{16 b(r)^2}{r^2} , \qquad \dot \psi = -f(r) \dot t \, . 
\end{equation}
These geodesics are important for understanding the causal structure of the spacetime.

\subsection{Causal structure and light cones}

As we have now seen, the soliton spacetime is relatively simple, horizon free, yet possesses a number of intriguing properties such as ergoregions. Below we will consider these spacetimes in view of the holographic complexity conjectures as they provide relatively simple examples of spacetimes with nontrivial angular momentum. An important aspect to understand for the application of the ``complexity equals action'' proposal is the causal structure of the spacetime. Morally, we expect the causal structure to be qualitatively the same as that of global AdS$_5$, and indeed we will see below that this expectation is borne out. Nonetheless, the causal structure of spacetimes with nontrivial angular momentum is far subtler than that for static geometries. Our purpose here is to discuss aspects of the causal structure for the solitons to make clear exactly why certain simplifications occur.

The complexity equals action conjecture asserts that the complexity of a state in the CFT and time $t$ is dual to the gravitational action evaluated on the Wheeler-DeWitt patch. This patch is defined to be the domain of dependence of a bulk Cauchy slice that asymptotically approaches the boundary time slice of interest. Naturally, this region is bounded by null hypersurfaces. In the context of rotating black holes, a careful treatment of null hypersurfaces was given in~\cite{Pretorius:1998sf}.\footnote{This analysis was recently extended to the asymptotically AdS case in~\cite{Balushi:2019pvr}.} Unlike the case for static solutions, in rotating cases null hypersurfaces do not generically lie in the $(t,r)$ plane --- in other words, the ``tortoise coordinate'' is a function not only of $r$, but is also depends nontrivally  on the angular direction(s). We will repeat the analysis of~\cite{Pretorius:1998sf} here and see that a dramatic simplification occurs in this case. Such foliations should find application also in the context of the propagation of wavefronts on these backgrounds in the high frequency limit.


Introduce a tortoise coordinate $r^* = r^*(r,\theta,\phi,\psi)$ and an associated in-going coordinate  $v = t + r^*$.  Demanding that the surfaces of constant $v$ be null  hypersurfaces requires $g^{-1}(\td v, \td v) =0$. A direct computation shows that it is possible to construct an additively separable solution of the form
\begin{equation}
r^* = R(r) + \Theta(\theta) + \Phi(\phi) + \Psi(\psi),
\end{equation} and by an appropriate choice of integration constants one can always arrange so that $r^* = r^*(r)$. This effectively amounts to choosing $SU(2) \times U(1)$-invariant hypersurfaces, so that $(\theta, \phi,\psi)$ are natural coordinates on spatial sections.   We have
\begin{equation}
\left(\frac{\td r^*}{\td r}\right)^2 = \frac{4 b^2(r)}{r^2 W(r)^2} \, ,
\end{equation}  
and we take the positive root. A completely analogous computation follows for the outgoing coordinate and we have
\begin{equation}
v \equiv t + r^*(r), \qquad u \equiv t - r^*(r) \, .
\end{equation} Surfaces of constant $u$ and $v$ are null hypersurfaces and the associated normal vector fields $g^{-1}(\td u, ~)$ and $g^{-1} (\td v, ~)$  are tangent to the (affinely) parameterized null geodesic generators of these null hypersurfaces.  A simple computation shows that the normals to these surfaces are proportional to the radial null geodesics presented in the previous section.

In the $(v,r^*)$ chart, the metric takes the form
\begin{equation}
g = \frac{r^2 W(r)}{4 b(r)^2} \left[ -\td v^2 + 2\td v \td r^* \right] + \frac{r^2}{4} (\td \theta^2 + \sin^2\theta \td \phi^2) + b(r)^2 \left( \td \psi + \cos\theta \td \phi + f(r) (\td v - \td r^*) \right)^2
\end{equation} 
with an analogous expression for the metric in the $(u,r^*)$ chart. It is is also convenient to write the metric in the double null coordinates
\begin{align}\label{solDubNull}
g &= \frac{r^2 W(r)}{4 b(r)^2} \left[ -\td u \td v\right] +  \frac{r^2}{4} (\td \theta^2 + \sin^2\theta \td \phi^2) 
\nonumber\\
&+ b(r)^2 \left( \td \psi + \cos\theta \td \phi +\frac{ f(r)}{2} (\td v + \td u) \right)^2 \, .
\end{align} In this form it is easy to read off the induced metric on the degenerate four-dimensional null hypersurfaces. 

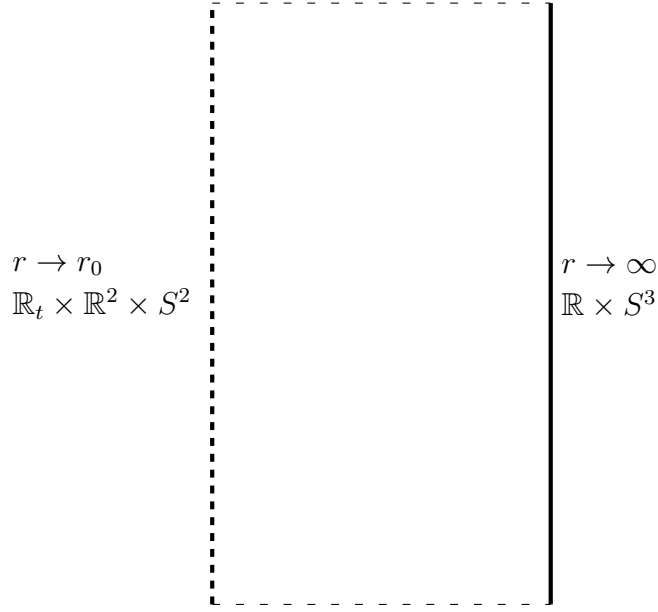
\begin{figure}
    \centering
    \begin{tikzpicture}

\draw[black,loosely dashed] (1,0) to (5.5,0);
\draw[black,loosely dashed] (1,8) to (5.5,8);












\draw[black,ultra thick] (5.5,0) to (5.5,8);
\node [ right, black] at (5.5,4.5) {$r \to \infty$};
\node [right, black] at (5.5,4) {$ \mathbb{R} \times S^3$};

\draw[black, dashed,ultra thick] (1,0) to (1,8);
\node [right, black] at (-1.8,4) {$\mathbb{R}_t \times \mathbb{R}^2 \times S^2$};
\node [right, black] at (-1.8,4.5) {$r \to r_0$};

\end{tikzpicture}

\caption{Causal structure of an asymptotically globally AdS$_5$ soliton. Spacetime in a neighbourhood of the `origin' (indicated by the dashed line on the left)  has topology $\mathbb{R} \times \mathbb{R}^2 \times S^2$.}    
    
\end{figure}

Let us investigate the function $r^*(r)$. As $r \to \infty$ one finds that by choosing the integration constant appropriately, $r^* =\ell \pi / 2 - \ell^2/r + \ell^4/(3r^3) + O(r^{-4})$, and hence  has the same asymptotic behaviour as $\ell \arctan(r/\ell)$, which is what one obtains in the pure AdS$_5$ case.  On the other hand, as $r \to r_0$, 
\begin{equation}
r^* = \left[\frac{2 j x}{(2+x)(1 - x(1-\alpha))^2}\right]^{1/2} (r-r_0)^{1/2} + O((r-r_0)^{3/2})
\end{equation} and, since 
\begin{equation}
b(r)^2 = \frac{1}{2} j \sqrt{x} (2+x)(r-r_0) + O(r-r_0)^2, \qquad f(r) = -\frac{2 x}{j(2+x)} + O(r-r_0),
\end{equation} we find that as $r \to r_0$
\begin{equation} 
\begin{aligned}
\td s^2 & \to \frac{r_0^2}{4}  ( \td \theta^2 + \sin^2\theta \td\phi^2) + (1 -x(1 - \alpha))\bigg[ -\td t^2 + (\td r^*)^2 
\\
&+ \frac{(2+x)^2(1 -x(1 - \alpha))(r^*)^2}{4}  \left( \td  \psi + \cos \theta \td \phi -\frac{2x}{j(2+x)} \td t\right)^2 \bigg]  \, . 
\end{aligned}
\end{equation} Using the regularity condition \eqref{reg} we find it is natural to interpret $r^*$ as a polar radial coordinate on the patch of $\mathbb{R}^2$ covered by the chart $(r^*,\psi)$ with origin at $r^* =0$.  In this chart is is clear that spacetime in a neighbourhood of the soliton is $\mathbb{R} \times \mathbb{R}^2 \times S^2$.

\section{Soliton mechanics}
\label{solMech}

Let us now turn to the construction of mass and mass variation formulas for the solitons.  In the asymptotically flat setting $(\ell \to \infty)$ the standard definitions of ADM mass and angular momenta can be applied.   As proved in~\cite{Kunduri:2013vka, Gunasekaran:2016nep}, the Smarr relation and first law in this case reads
 \begin{equation}\label{AFmech}
 M = \frac{\Psi[C] q[C]}{2}, \qquad \td M = \Psi[C] \td q[C] \, ,
 \end{equation} where
 \begin{equation}
 \Psi[C] = \pi U_{\bar \psi}(r_0) = \frac{\sqrt{3} \pi r_0^2 (j^2 + r_0^2)}{j^3}
 \end{equation} is a thermodynamic potential associated to the 2-cycle.  Here the first law should be understood as a relationship between the physical parameters in the space of smooth solutions. In particular, this means that the validity of the first law is equivalent to the regularity condition~\eqref{reg} (with $\alpha = 0$).
 
 The derivation of these identities relied on an application of Stokes' theorem to the Komar definition of the mass and first variation of the mass.  As is well known, in the presence of a negative cosmological constant this gives a divergent expression for the mass.  Nonetheless it is natural to expect that there exists a suitable generalization of \eqref{AFmech}. We will show this is indeed the case by revisiting the approach used by Kastor, Ray, and Traschen in the context of black holes~\cite{Kastor:2009wy}. This requires using a suitably regularized Komar integral and the Ashtekar-Magnon definition of the mass of a stationary, asymptotically globally AdS spacetime. 
 
 To begin, note that the stationary Killing field $\xi$ is automatically divergenceless, i.e. $\td \star \xi=0$. Thus the four-form $\star \xi$ is closed, although it need not be exact.  We may introduce a locally defined 2-form `Killing potential' $\omega$ satisfying
 \begin{equation}
 \star \xi = -\td \star \omega
 \end{equation} or equivalently $\xi = \star \td \star \omega$.  Clearly there is a gauge freedom in choosing $\omega$ which we will fix below.  The motivation for the definition arises from the identity
 \begin{equation}
 \td \star \td \xi = 2 \star \text{Ric}(\xi) = \frac{8}{\ell^2} \td \star \omega + 2 \star \text{T}_{EM}(\xi)
 \end{equation} where $\text{T}_{EM}$ refers to the contribution to the Ricci tensor arising from the Maxwell field.  As shown in~\cite{Kunduri:2013vka}, 
 \begin{equation}
 \star \text{T}_{EM}(\xi) = -\frac{1}{3} \Theta \wedge F + \frac{4}{3} \td (\star F \Phi_\xi).
 \end{equation} Let $S^3_\infty$ stand for asymptotic sphere at infinity at constant time.  We then have, assuming $\star \omega$ is globally defined on a spacelike hypersurface $\Sigma$
 \begin{equation}\label{Stokes}
 \int_{\Sigma} \td \left(\star \td \xi - \frac{8}{\ell^2} \star \omega \right) = \int_{S^3_\infty} \left(\star \td \xi - \frac{8}{\ell^2} \star \omega\right)  - 
 \int_{\partial\Sigma_{int}} \left(\star \td \xi - \frac{8}{\ell^2} \star \omega\right) \, , 
 \end{equation} where we have included a possible interior boundary $\partial \Sigma_{int}$.   Note that there is no horizon in the case of gravitational solitons and hence there are no horizon contributions to $\partial \Sigma_{int}$.  For the Killing potential $\omega$ we derive below, it  will  be appropriate to choose $\Sigma$ to have a boundary at some $r = r_0 + \epsilon$, $\epsilon > 0$ and then take $\epsilon \to 0$.  In this limit, $\Sigma$ will be a complete hypersurface that smoothly degenerates at $r = r_0$ and extends to spatial infinity. The key observation of~\cite{Kastor:2009wy, Cvetic:2010jb} was that the $S^3_\infty$ boundary integral is  finite and with a suitable gauge choice for $\omega$, can be set to agree with  the Ashtekar-Magnon mass.  
 
Now let us explicitly construct the Killing potential $\omega$. It is useful to use the orthonormal basis
 \begin{align}
e^0  &= \frac{r \sqrt{W(r)}}{2 b(r)} \td t ,  \qquad e^1 = \frac{ \td r}{\sqrt{W(r)}}, \quad e^2 = b(r) (\td \psi + \cos\theta \td \phi + f(r) \td t), \\ e^3 &= \frac{r}{2} \td \theta, \quad e^4 = \frac{r}{2} \sin \theta \td \phi \, ,
\end{align} in which the stationary Killing field is written
\begin{equation}
\xi = -\frac{ r \sqrt{W}}{2 b(r)} e^0  + b(r) f(r)e^2 \, .
\end{equation} The geometry is cohomogeneity-one and hence it suggests we seek a solution for $\omega$ of the form $\omega = \tfrac{1}{2} \omega_{ab} (r) e^a \wedge e^b$.  We omit the details of the computation and simply give a solution
\begin{equation}
\begin{aligned}
\omega &= \left( \frac{r^4 - 8 C_1}{16 r b(r)^2} - \frac{b(r)^2 f(r)^2 (r^4 - 8 C_1)}{4 r^3 W(r)} + \frac{b(r)^2 f(r) C_2}{r^3 W(r)}\right) \td t \wedge \td r \\ &+ \frac{b(r)^2}{W(r)}\left(\frac{f(r) r}{4} - \frac{2 C_1 f(r)}{r^3} - \frac{C_2}{r^3}\right) \td r \wedge \sigma_3
\end{aligned}
\end{equation} with Hodge dual
\begin{equation}
\star \omega = -\frac{(r^4 - 8 C_1)}{32} \sin\theta \td \psi \wedge \td \theta \wedge \td \phi - \frac{C_2}{8} \sin\theta \td t \wedge \td \theta \wedge \td \phi.
\end{equation} 
By inspection, note that because $i_{\partial_\psi}\star \omega \neq 0$, $\omega$ will not in general be regular at the `origin' $r = r_0$ unless we also fix $C_1 = r_0^4/8$ (this can be explicitly checked by passing to a Cartesian coordinate system centered at $r = r_0$).  However, for our purposes it will be more natural to choose $C_1 $ so that the first integral appearing on the right hand side of \eqref{Stokes} is the Ashtekar-Magnon mass $M$:\footnote{Note that if the first choice was made --- i.e. to ensure regularity of the Killing potential at the origin --- this could allow one to interpret the thermodynamic volume as a boundary quantity. However, while there may be advantages to this approach worth exploring, here we follow the conventional method of picking the gauge of the Killing potential to yield the correct mass.}
\begin{equation}
M = \frac{\pi}{4} \left(3(p-q) + \frac{j^2 p}{\ell^2}\right) \, .
\end{equation} We have
\begin{equation}
I_1 \equiv \int_{S^3_\infty} \left(\star \td \xi - \frac{8}{\ell^2} \star \omega \right) = 8\pi^2 \left(q - p  - \frac{4 C_1 + j^2 p }{\ell^2}\right) \, ,
\end{equation} therefore if we fix $C_1 = -p j^2/6$, we find
\begin{equation}
M = -\frac{3}{32\pi} I_1.
\end{equation}  We now apply the identity \eqref{Stokes} to a spacelike hypersurface $\Sigma$ with asymptotic boundary $S^3_\infty$ and  an interior boundary at $r = r_0 + \epsilon$. Taking $\epsilon \to 0$ we find 
\begin{equation}
\int_{\partial \Sigma_{int}} \star \td \xi \to 0 \, ,
\end{equation} whereas we identify the thermodynamic volume as in~\cite{Kastor:2009wy} by
\begin{equation}\label{TDvol}
V \equiv  - \int_{\partial \Sigma_{int}} \star \omega \to \frac{r_0^4 \pi^2}{2}  - 4C_1 \pi^2= \frac{\pi^2 r_0^4}{6} \left( 1 + \frac{2 r_0^2}{j^2}\right) \, .
\end{equation}  
Note that the first term is a `naive' geometric volume in the sense that it could be interpreted as the volume of a four-dimensional ball of radius $r_0$ with volume form associated to the induced metric $g$ on $\Sigma$:
\begin{equation}
\frac{r_0^4 \pi^2}{2} = \int_0^{r_0} \int_{S^3} \td \text{Vol}(g).
\end{equation} Intuitively this represents the volume of the part of AdS$_5$ that has been `removed' by the presence of the soliton.

Following the standard prescription, we define the thermodynamic pressure of AdS$_5$ 
\begin{equation}
P \equiv -\frac{\Lambda}{8 \pi G} = \frac{3}{4\pi \ell^2}.
\end{equation} We then have the Smarr-type relation
\begin{equation}
M + PV = -\frac{3}{32\pi}\int_{\Sigma} \left( \td \star \td \xi - \frac{8}{\ell^2} \td \star \omega \right) = \frac{1}{16\pi} \int_\Sigma \left( \Theta \wedge F - 4 \td (\Phi_\xi \star F) \right)  \, .
\end{equation} We can discard the exact term by working in a gauge where $\Phi_\xi \to 0$ as $r \to \infty$ --- see \eqref{Max:Phi}. Therefore we need only compute the volume integral of $\Theta \wedge F$. This has been worked out for a general rod structure in the asymptotically flat setting in~\cite{Kunduri:2013vka}, although that analysis does not directly carry to the AdS setting. This is because in the asymptotically flat case, the potentials $U_i$ can be chosen to vanish at infinity. However, as is clear from \eqref{Max:U}, this cannot be arranged in the AdS case and so we will find additional terms. 

Evaluating the volume integral, 
\begin{equation}
\frac{1}{16\pi} \int_{\Sigma} \Theta \wedge F = \frac{\pi}{4}\int_{\mathcal{B}} \eta_{ij} \td U_j \wedge \td \Phi_i = \frac{\pi}{4} \int_{\mathcal{B}} \td [ \eta_{ij} U_j \td \Phi_i ] = \frac{\pi}{4} \int_{\partial \mathcal{B}} \eta_{ij} U_j \td \Phi_i \, .
\end{equation} Note that since $\td \Phi_i \to 0 $ as spatial infinity, there are no contributions from the asymptotic boundary $\partial\mathcal{B}_\infty$ of the orbit space.  Recall now that $\partial \mathcal{B}$ consists of intervals (`rods') corresponding  to the fixed point sets of a particular integer linear combination of Killing fields with $2\pi-$periodic orbits.   A finite rod corresponds to a compact 2-cycle ($S^2$) in the spacetime. For an asymptotically globally AdS or Minkowski spacetime, there will be two semi-infinite rods corresponding to asymptotic axes of symmetry (these correspond to non-compact discs). As explained in~\cite{Kunduri:2013vka}, we can write
\begin{equation}
\int_{\partial \mathcal{B}} \eta_{ij} U_j \td \Phi_i = \sum_{[C]} \frac{2}{\pi}\Psi[C] q[C]
\end{equation} where $q[C]$ is the dipole flux associated to a 2-cycle and $\Psi[C] = \pi v^i U_i$ is a (constant) potential associated to a cycle (or infinite disc) $C$ and $v^i \in \mathbb{Z}$ are the components of the `rod vector' which specifies the vanishing Killing field.  In contrast to the asymptotically flat case, we now find that the semi-infinite rods also contribute to this sum.  In terms of standard basis of Killing fields that are orthogonal at spatial infinity, 
\begin{equation}
\frac{\partial}{\partial \phi^1} = \frac{1}{2} \frac{\partial}{\partial \hat \psi} - \frac{\partial}{\partial \phi}, \qquad \frac{\partial}{\partial \phi^2} = \frac{1}{2} \frac{\partial}{\partial \hat \psi} + \frac{\partial}{\partial \phi}, 
\end{equation} the rod intervals and vectors of the gravitational soliton are (i) a semi-infinite rod $I_+$ with direction  $(1,0)$ corresponding to the set $\theta =0, r \in (r_0, \infty)$; a finite rod $I_C$ with direction $(1,1)$ with $r= r_0, \theta \in (0,\pi)$, and finally a semi-infinite rod $I_-$ with direction $(0,1)$ with $\theta = \pi, r \in (r_0, \infty)$. We have already given $\Psi[C]$ and $q[C]$ above.  We also find
\begin{equation}
\Psi[I_+] =\Psi[I_-] =  -\frac{\sqrt{3} \pi r_0^4}{2 j \ell^2} 
\end{equation} with the associated fluxes
\begin{equation}
q[D_+] = q[D_-] = -\frac{\sqrt{3} r_0^2}{4 j}.
\end{equation} 
in this case one finds that generalized Smarr identity 
\begin{equation}
M + P V = \frac{1}{2} \sum_{[C]} \Psi[C] q[C]
\end{equation} does indeed hold. 

We also expect a `first law' of soliton mechanics to hold, expressing an identity between variations in the phase space of solutions.  The parameters of the solution are $(r_0, j, \ell)$. However, not all of these parameters are independent --- they must satisfy the regularity condition~\eqref{reg}. 
Variations in the space of regular solutions must respect this condition. It follows that if we define
\begin{equation}
\mathcal{R} = 3 - \frac{3 r_0^4}{j^4} - \frac{r_0^6}{j^6}  + \frac{r_0^2(2 j^2 + r_0^2)^2}{j^4 \ell^2} = 0, 
\end{equation} then we must have $\td \mathcal{R} =0$ for all variations $(\td r_0, \td j , \td \ell)$.  A calculation shows that
\begin{equation}
\td M = V \td P + \Psi[C] \td q[C] + \Psi[D_+] \td q[D_+] + \Phi [D_-] \td q[D_-] + \frac{\pi j^2 r_0^2}{8(2j^2 + r_0^2)} \td \mathcal{R}
\end{equation} and hence we have the first law
\begin{equation}
\td M = V \td P + \sum_{[I]} \Psi[I] \td q[I].
\end{equation} This is a natural extension of the first law of soliton mechanics derived in the asymptotically flat setting. 

Note that, at first glance, the first law as written above appears to have more parameters varied than there are independent parameters of the solution. However, this is not the case. In this setting the $q[I]$ cannot be independently varied --- in fact, from the above expressions it can be seen that all $q[I]$ are identical (up to signs). In this sense the first law is not sensitive to the geometric origin of the different potentials $\Psi[I]$, as the variation of the mass at fixed $P$ yields only the sum of the potentials. It could be interesting to understand the first law in the setting where the three $U(1)$ charges are independent, e.g. the soliton solutions to $U(1)^3$ gauged supergravity discussed in~\cite{Cvetic:2005zi}.

Finally, it is worth mentioning that while the solitons have nontrivial thermodynamic volume their ``area'' is vanishing. This is interesting in light of the `reverse' isoperimetric inequality proposed in~\cite{Cvetic:2010jb}. This inequality, proposed for black holes, asserts that the isoperimetric ratio of the thermodynamic volume to area should satisfy
\be 
\mathscr{R} \equiv \left[\frac{V}{(D-1) \Omega_{D-2}} \right]^{1/(D-1)} \left[\frac{ \Omega_{D-2}}{A} \right]^{1/(D-1)} \ge 1 \, .
\ee
In this case it seems that there are two ways to interpret the result of trivial area but non-vanishing thermodynamic volume. One could argue that the inequality as initially proposed should apply only in the case of black holes, in which case our results would fall outside of its scope. Another option would be to view our case as `trivially' satisfying the inequality, i.e. with $\mathscr{R} \to \infty$. From this latter perspective, one could interpret the reverse isoperimetric inequality as the physical statement that {\it a spacetime cannot possess entropy without thermodynamic volume}.

\section{Holographic complexity}
\label{holoComp}

We now consider the soliton spacetime in light of the holographic complexity conjectures. Our motivation here is simple: The soliton spacetime represents a relatively simple spacetime possessing angular momentum, with a straightforward causal structure. To date, most studies of holographic complexity have focused on solutions without rotation. It is our aim here to perform some prelimary investigations of the effect of rotation and to determine what --- if any ---  effects is has in this circumstance. The soliton spacetime does not possess an `interior' and therefore we do not expect time dependence in the action. Nonetheless, we are able to compute the holographic `complexity of formation'~\cite{Chapman:2016hwi} for the soliton using global AdS as the reference state.  Here we will consider both the ``complexity equals volume'' and ``complexity equals action'' conjectures.

\subsection{Complexity equals volume}

The complexity equals volume (CV) conjecture asserts that the complexity of the state of the boundary CFT is dual to the extremal codimension-one bulk hypersurface that meets the boundary on the timeslice where that state is defined~\cite{Stanford:2014jda}. For the soliton, surfaces of constant $t$ are extremal slices. The complexity of formation within the CV proposal can then be obtained by determining the volume of the $t = 0$ hypersurface and subtracting the corresponding result for global AdS. A simple computation reveals that the volume of the constant $t$ surface can be obtained as the following integral
\be 
{\cal V} = 4 \pi^2 \int_{r_0}^{\rmax} \frac{r^2 b(r)}{\sqrt{W(r)}} \td r \, .
\ee
The expression for global AdS is qualitatively similar with the appropriate substitutions made for the metric functions:
\be 
{\cal V}_{\rm AdS} = 2 \pi^2 \int_{0}^{\rmax} \frac{r^3}{\sqrt{1 + r^2/\ell^2}} \td r \, .
\ee
According to the CV prescription, the complexity of formation is given by
\be 
\Delta \mathcal{C}_\mathcal{V} = \frac{\Delta \mathcal{V}}{G L}
\ee
where $L$ is some (arbitrary) length scale --- here, consistent with~\cite{Chapman:2016hwi}, we will take $L = \ell$.

\begin{figure}[htp]
\centering
\includegraphics[width=0.43\textwidth]{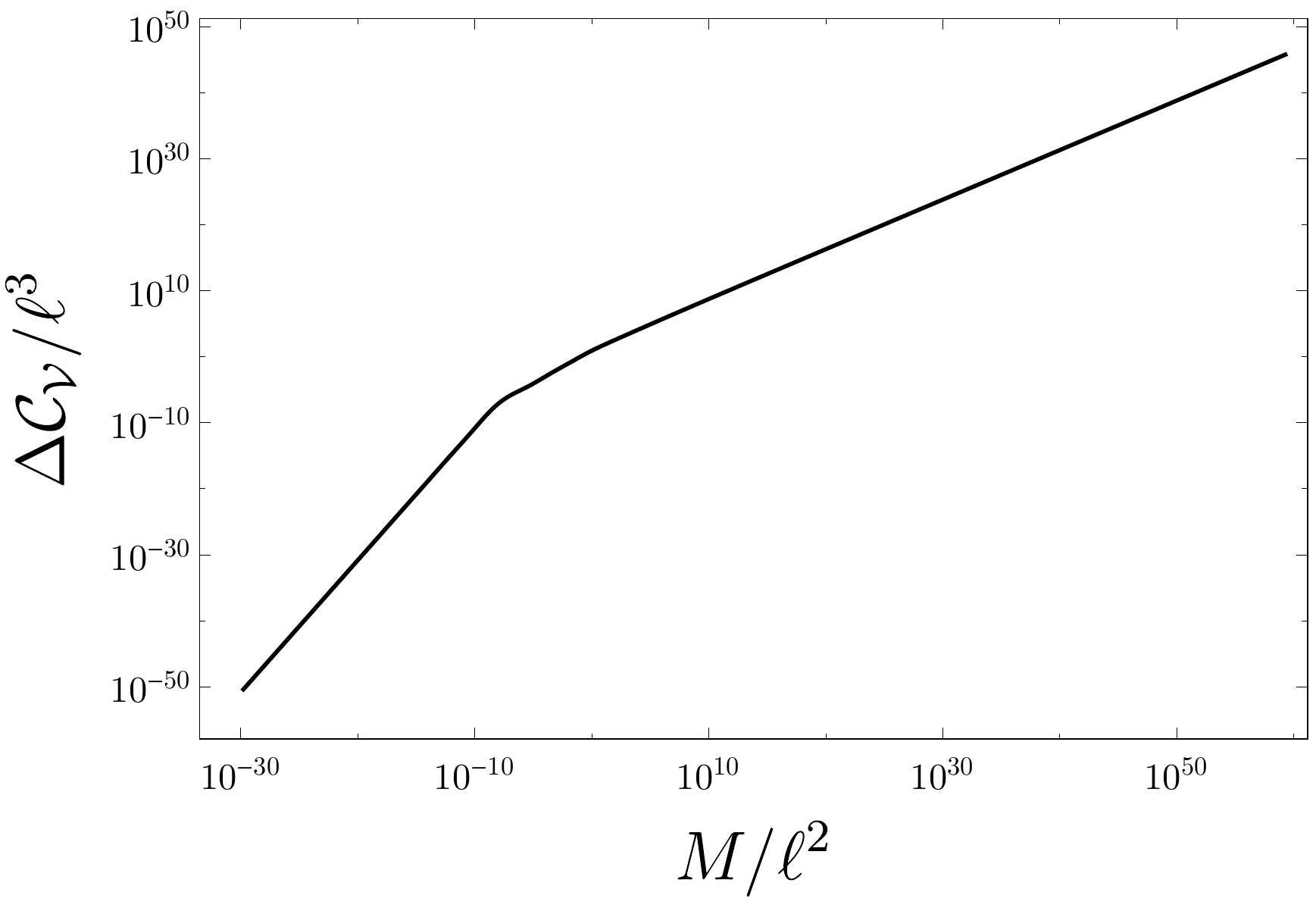}
\quad
\includegraphics[width=0.43\textwidth]{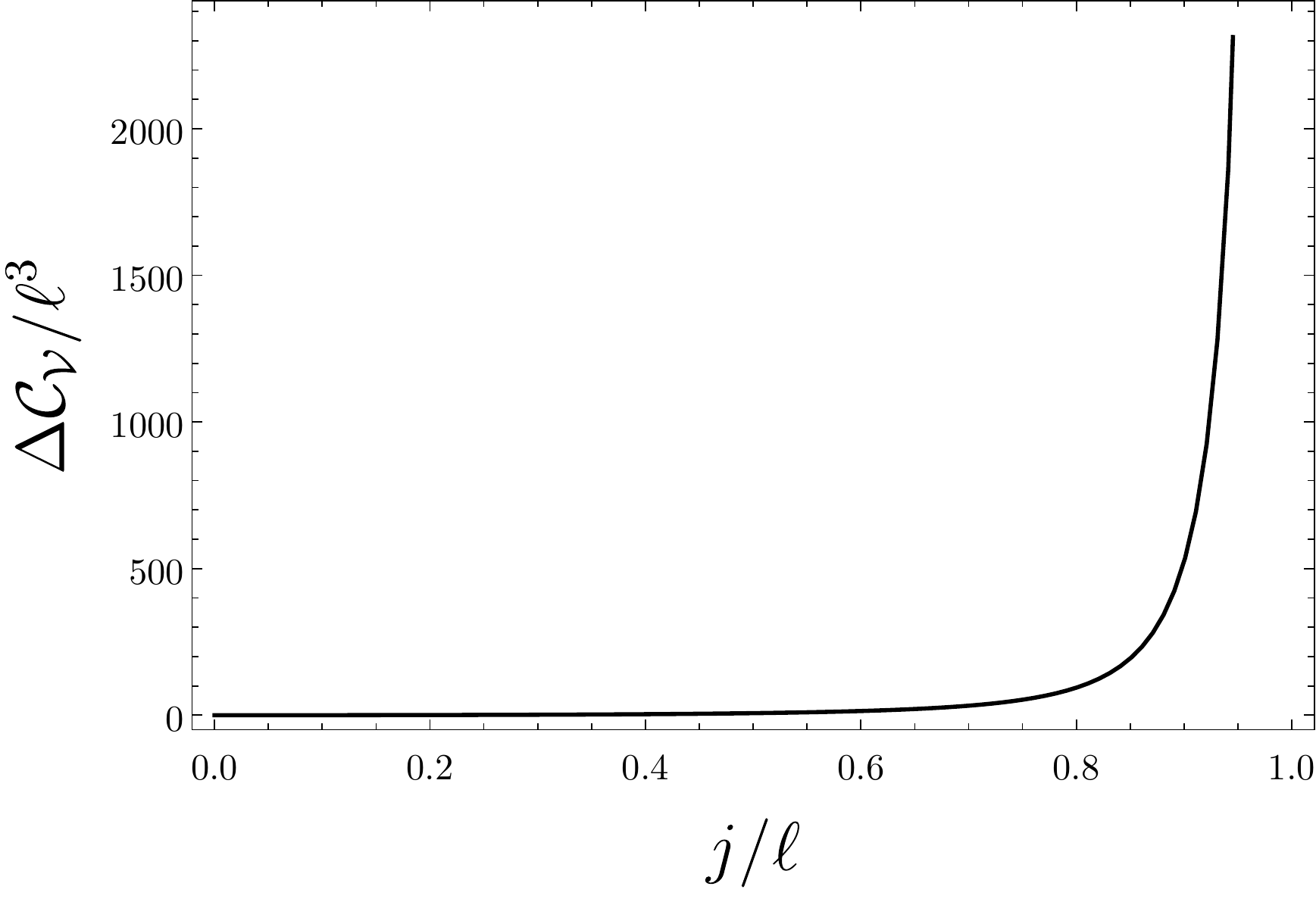}
\caption{Left: A log-log plot of the complexity of formation for the CV proposal plotted as a function of the mass. The complexity of formation is a monotonically increasing function of the mass.  Right: Here $\Delta \mathcal{C}_\mathcal{V}$ is plotted against $j/\ell$ to make more direct the comparison with the action later. This plot was produced for units in which $\ell = 1$.}
\label{CVform}
\end{figure}

It is not possible to find a useful closed form expression for the complexity of formation.  Nonetheless, the complexity of formation can be easily evaluated numerically and in figure~\ref{CVform} we show an illustrative plot. The complexity of formation is always positive and is a monotonically increasing function of the black hole mass, diverging to $+\infty$ in the limit $j\to \ell$. On the left, we display the complexity of formation on log-log scale as it makes manifest the two asymptotic regimes for which the dependence is rational power of the mass. Let us discuss these regimes in more detail. 

We can make analytic progress in the limit $y \equiv r_0/\ell  \ll 1$ (see appendix~\ref{smallSol} for additional details). In this case, the integrals can be evaluated perturbatively with the result
\be 
\Delta \mathcal{C}_\mathcal{V} = \frac{(1 + j_0^2)\pi^2 \ell^3}{j_0^4} y^2 + {\cal O}(y^3) \, .
\ee
This leading-order expression can be usefully compared with the analogous one for the mass:
\be 
M = \frac{(1 + j_0^2)\pi^2 \ell^4}{j_0^4} y^2 + {\cal O}(y^3) \, .
\ee
Thus, to leading order we have
\be 
\Delta \mathcal{C}_\mathcal{V}  \approx \frac{8 \pi }{3 \ell} M \, . 
\ee
Note that this proportionality holds only to leading order --- at higher order in $y$ a more complicated functional form arises. Nonetheless, it is somewhat pleasing to see that the leading-order deviation from the vacuum is proportional to the mass. Within the framework of black hole chemistry, the mass should be interpreted as spacetime enthalpy. Recall that the enthalpy corresponds to the energy of a system plus the amount of work required to place the system in its environment. This provides a nice thermodynamic interpretation of the complexity of formation for small solitons: it is the ``enthalpy of formation'' of the spacetime.

It is also useful to study the behaviour of $\Delta \mathcal{C}_\mathcal{V}$ for large $r_0$. In this regime we work numerically to extract information about the structure of the integral. By constructing a logarithmic plot of the complexity of formation, we can extract that $\Delta \mathcal{C}_\mathcal{V}$ behaves in the following way at large $r_0$:
\be
\Delta \mathcal{C}_\mathcal{V} \approx  24.398 \frac{r_0^{9/2}}{\ell^{3/2}} + \text{subleading} \, ,
\ee
where the prefactor was determined numerically to the precision shown. Again, we could compare this behaviour with the mass of the solution, concluding that 
\be 
\Delta \mathcal{C}_\mathcal{V} \propto \ell^{3/2} M^{3/4} \, .
\ee
However, it is also fruitful to compare this behaviour with the thermodynamic volume. In fact, in the large $r_0$ limit, the thermodynamic volume --- see eq.~\eqref{TDvol}} --- behaves as
\be 
V = \frac{\pi^2 r_0^6}{3 \ell^2} + \cdots 
\ee
and therefore the behaviour is precisely the following:
\be 
\Delta \mathcal{C}_\mathcal{V} \approx 9.988 V^{3/4} + \text{subleading} \, .
\ee

It is worth remarking that the functional form of the complexity formation, being proportional to $V^{3/4} (= V^{(D-2)/(D-1)})$, is true also for large static black holes. As considered in~\cite{Chapman:2016hwi}, the complexity of formation for large (static) black holes is proportional to the entropy with a dimensionless constant of proportionality. Recall~\cite{Gunasekaran:2012dq} that for $D$-dimensional static black holes we have $S \propto V^{(D-2)/(D-1)}$, and therefore the same functional form holds in both cases. We emphasize, however, that the soliton is horizonless and therefore has vanishing entropy. It is therefore somewhat interesting that these results coincide to the extent that they do. It could be worthwhile to explore this connection further, perhaps in the context of rotating black holes.

\subsection{Complexity equals action}

The ``complexity equals action'' conjecture asserts that the complexity of the boundary CFT state is given by the value of the gravitational action evaluated on the Wheeler-DeWitt (WDW) patch~\cite{Brown:2015bva, Brown:2015lvg}:
\be 
\mathcal{C}_\mathcal{A} = \frac{I_{\rm WDW}}{\pi} \, .
\ee
The WDW patch is the region of spacetime bounded by past and future lightsheets that meet the boundary at the timeslice where the CFT state is defined.

The WDW patch is a nontrivial spacetime region, including light-like boundaries and nonsmooth junctions of these surfaces. A careful treatment of the gravitational action for such regions was provided in~\cite{Lehner:2016vdi}.\footnote{See also \cite{Booth:2001gx, Parattu:2015gga, Cano:2018ckq, Jiang:2018sqj} for previous work and generalizations of this formalism.} Including all relevant boundary, joint, and counterterm contributions in the action we have 
\begin{align}
I =& \, I_{\rm bulk} + \frac{1}{8 \pi G} \int_{\mathcal{B}} \td^4 x \sqrt{|h|} K + \frac{1}{8 \pi G} \int_{\mathcal{B}'} \td \lambda \td^3 x \sqrt{\gamma} \kappa  + \frac{1}{8 \pi G} \int_{\Sigma'} \td^3 x \sqrt{\sigma} a  
\nonumber\\
&+ \frac{1}{8 \pi G} \int_{\mathcal{B}'} \td \lambda \td^3 x \sqrt{\gamma} \Theta \log \left( L_{\rm ct} \Theta \right)  \, .
\end{align}
We use the conventions of~\cite{Carmi:2016wjl} (c.f. appendix A of that work),\footnote{Note that a typo in the sign of the null boundary term was corrected in~\cite{Chapman:2018dem}. We have included this correction here, but note that it will have no implications on our results since the null generators will be affinely parameterized.} and refer the reader there for a detailed discussion of these conventions. Here let us just note some of the more relevant details.

The first term appearing here is the bulk action as given in eq.~\eqref{minsugra} including both the usual Einstein-Hilbert and the Maxwell-Chern-Simons contributions. Besides the bulk term, the next two terms appearing in the full action are the surface terms.  The first is the usual Gibbons-Hawking-York (GHY) boundary term for spacelike/timelike surfaces. The convention here is that the normal \textit{one-form} is always outward pointing. With this convention, the GHY term as written applies equally well for spacelike/timelike boundaries. The second term above is the surface term for null boundaries. For a segment of null boundary with normal $k^\alpha$, $\kappa$ is defined in the usual way: $k^\beta \nabla_\beta k^\alpha = \kappa  k^\alpha$, while $\gamma$ is the determinant of the metric on the  $(D-2)$-dimensional cross-sections of the null boundary and $\lambda$ is defined by $ k^\alpha = \partial x^\alpha / \partial \lambda$.

The fourth term appearing in the action is the joint contribution arising for the intersection of  null and timelike/spacelike boundaries. The parameter $a$ is defined according to
\begin{align}
\text{timelike/null}: \quad a & \equiv \epsilon \log | \mathbf{t}_1 \cdot \mathbf{k}_2 | \quad \text{with} \quad \epsilon = - {\rm sign}(\mathbf{t}_1 \cdot \mathbf{k}_2) {\rm  sign} (\hat{\mathbf{n}}_1 \cdot \mathbf{k}_2 ) \, ,
\\
\text{null/spacelike}: \quad a & \equiv \epsilon \log | \mathbf{k}_1 \cdot \mathbf{n}_2 | \quad \text{with} \quad \epsilon = - {\rm sign}(\mathbf{k}_1 \cdot \mathbf{n}_2) {\rm  sign} (\mathbf{k}_1 \cdot \hat{\mathbf{t}}_2 ) \, ,
\\
\text{null/null}: \quad a & \equiv \epsilon \log | \mathbf{k}_1 \cdot \mathbf{k}_2 | \quad \text{with} \quad \epsilon = - {\rm sign}(\mathbf{k}_1 \cdot \mathbf{k}_2) {\rm  sign} (\hat{\mathbf{k}}_1 \cdot \mathbf{k}_2 ) \, ,
\end{align}
where $\mathbf{k}_i$ is a null normal, $\mathbf{t}_i$ is a timelike unit normal, and $\mathbf{n}_i$ is a spacelike unit normal. Additionally, depending on the intersecting boundary segments, auxillary vectors --- indicated with a hat --- are required. These unit vectors are defined by the conditions of living in the tangent space of the appropriate boundary segment and pointing outward \textit{as a vector} from the joint of interest.

The final term in the action is a counterterm for the null boundaries. Unlike the other terms just discussed, this quantity is not required to have a well-posed variational principle. Rather this counterterm ensures the invariance of the action under reparameterizations of the generators of the null boundaries. While this term was not explicitly considered in the initial investigations of complexity of formation (indeed, its contribution in this context for static black holes vanishes), it plays an essential role for reproducing desired properties of complexity (e.g. the switchback effect) in more complicated scenarios~\cite{Susskind:2014jwa,  Chapman:2018dem, Chapman:2018lsv}. Note that the counterterm depends on an arbitrary length scale $L_{ct}$. Here we will consider the implications of this term for the complexity of formation of the solitons.

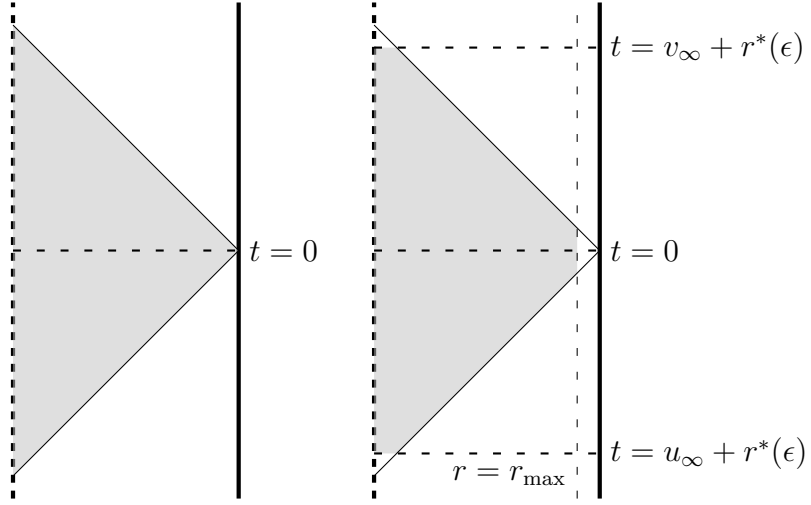
\begin{figure}[!ht]
    \centering
    \hspace{100pt}
    \begin{tikzpicture}[scale=0.6]













\draw[black,ultra thick] (3,-0.5) to (3,10.5);

\draw[black, dashed,ultra thick] (-2,-0.5) to (-2,10.5);

\fill[fill=lightgray,opacity=0.5]  (-2,0) -- (3,5) -- (-2,10);
\draw[black] (-2,0) -- (3,5) -- (-2,10);
\draw[black, loosely dashed, thick] (-2,5) to (3,5);
\node [right, black] at (3,5) {$ t =0$};


\draw[black,ultra thick] (11,-0.5) to (11,10.5);

\draw[black, dashed,ultra thick] (6,-0.5) to (6,10.5);

\fill[fill=lightgray,opacity=0.5]  (6,0.5) -- (6.5,0.5) -- (10.5,4.5) --(10.5,5.5) --( 6.5,9.5 ) -- (6,9.5);
\draw[black, loosely dashed] (10.5,-0.5) to (10.5,10.5);
\node [left,black] at (10.5,0) {$r = r_{\text{max}}$};
\draw[black] (6,0) -- (11,5) -- (6,10);
\draw[black, loosely dashed, thick] (6,5) to (11,5);
\node [right, black] at (11,5) {$ t =0$};

\draw[black,loosely dashed, thick] (6,9.5) to (11,9.5);
\node[right,black] at (11,9.5) {$t = v_\infty + r^*(\epsilon)$};

\draw[black,loosely dashed,thick] (6,0.5)--(11,0.5);
\node[right,black] at (11,0.5) {$t = u_\infty + r^*(\epsilon)$};


\end{tikzpicture}
\caption{Left: Wheeler-DeWitt patch is indicated in grey. Right: regulated Wheeler-DeWitt patch.}    
    \label{WDWpatch}
\end{figure}

As made clear in the earlier discussion, the causal structure of the soliton spacetime is qualitatively similar to that of global AdS itself. Therefore, the Wheeler-DeWitt patch is also qualitatively similar. We show the WDW patch and the regulated version in figure~\ref{WDWpatch}.  The patch is bounded by two null hypersurfaces which we will refer to as $P$ and $F$ that respectively denote the past and future null boundaries. The surface $F$ is a surface of constant $v$: $v_\infty = t + r^*(r)$, while $P$ is a surface of constant $u$: $u_\infty = t - r^*(r)$. These surfaces meet at the AdS boundary on the $t = 0$ hypersurface, a constraint that fixes the constants $v_\infty$ and $u_\infty$ to be
\be 
v_\infty = -u_\infty = \lim_{r \to \infty} r^*(r) \, .
\ee
In the computations below, it will be convenient to choose the integration constant in the tortoise coordinate such that $v_\infty$ and $u_\infty$ match those for the AdS vacuum. Explicitly, this amounts to setting
\be 
r^*(r) = \frac{\pi \ell}{2} + \int_\infty^r \frac{2 b(\tilde{r})}{\tilde{r} W(\tilde{r})} \td \tilde{r} \, .
\ee
In the bulk, the null boundaries of the WDW patch develop caustics at $r_0$ where the coordinate $\psi$ degenerates. However, just as in the vacuum case, the caustics at the future and past tips of the WDW patch make no contribution to the complexity of formation --- see appendix~\ref{furtherDeets}.

The action of the WDW patch is, of course, divergent and therefore we must regulate it by introducing a large distance cutoff at $r = r_{\rm max}$ prior to subtracting the contribution of AdS.  Note that we will use the same cutoff $\rmax$ for both the AdS and soliton geometries. This is justified by the results in appendix~\ref{FGstress}. Further, we will not present an explicit discussion of the GHY term at $\rmax$ in the main text --- as expected, this term cancels with the analogous one arising in the AdS computation, as we justify in appendix~\ref{furtherDeets}. Let us now consider each of the relevant contributions in turn.

\subsubsection{Evaluating the action}

\subsubsection*{Null boundaries}

First consider $F$ --- the future null boundary of the WDW patch. The outward-directed normal one-form to $F$ is
\be 
(k_F)_\mu \td x^\mu = \alpha \left[ \td t + \td r^* \right]
\ee
where $\alpha$ is an arbitrary, positive constant  (introduced to allow us to track the dependence of the parameterization of the null normals) and
\be 
\frac{\td r^*}{\td r} = \frac{2 b}{r W} \, .
\ee 
We can easily see that this null generator is affinely parameterized. Namely, $\td k_F = 0$ and so $\kappa = 0$ necessarily. To determine the affine parameter, first note that as a vector
\be 
(k_F)^\alpha \partial_\alpha =  \partial_\lambda = \alpha \left[ - \frac{4 b^2}{r^2 W} \partial_t + \frac{2 b}{r} \partial_r + \frac{4 b^2 f}{r^2 W} \partial_\psi \right] \, .
\ee
This is clearly proportional to the radial null geodesics constructed in section~\ref{ergoGeo}. As clear from that section, we can arrange things such that 
\be 
\td \lambda_F = \frac{r}{2 \alpha b} \td r \, ,
\ee
describes a one-parameter family of affine parameterizations. Note that in the limit where $j=0$ the solution reduces to global AdS and we have $b^2 = r^2/4$ yielding $\lambda_F = r/\alpha$ as the affine parameter.\footnote{This is consistent with appendix E of~\cite{Carmi:2017jqz}.} For the soliton the situation is a bit more complicated, and unfortunately it seems that it is not possible to express $\lambda_F(r)$ in a simple closed form. However, as we will see below, this will not be necessary.

Since the null generator is affinely parameterized, the null boundary term does not contribute. However, to ensure the parameterization independence of the action we must consider also the counterterm contribution. Explicitly, the counterterm contribution is 
\be 
I_{ct} = \frac{1}{8 \pi G} \int \td \lambda \td^3x \sqrt{\gamma} \Theta \log (L_{ct} \Theta)
\ee
where
\be 
\Theta = \partial_\lambda \log \sqrt{\gamma} 
\ee
is the expansion and $\gamma$ is the determinant of the metric on cross-sections of the null hypersurface. For the soliton metric, it is easy to read this off from the form of the metric presented in eq.~\eqref{solDubNull}:
\be 
\sqrt{\gamma} = \frac{r^2 b \sin\theta}{4} \, .
\ee
Therefore, we obtain the following expressions for the expansion
\begin{align} 
\Theta_F &= \frac{\td}{\td\lambda_F} \log \sqrt{\gamma} = \frac{1}{2 \gamma} \frac{\td}{\td\lambda_F} \gamma = \frac{1}{2 \gamma} \frac{\td r}{\td\lambda_F} \frac{\td}{\td r} \gamma \, , 
\\
&= \left[\frac{2}{r} + \frac{ (b^2)'}{2 b^2} \right] \frac{\td r}{\td\lambda_F} \, .
\end{align}
Note that upon substituting $j = 0$, we obtain $\Theta_F = 3 \alpha/r$, which is just as we would expect for the AdS vacuum.

Now, putting things together we have
\begin{align} 
I^F_{ct} &= \frac{1}{8 \pi G} \int \td\lambda \td^3x \sqrt{\gamma} \Theta \log (L_{ct} \Theta) \, ,
\\
&= \frac{\pi}{2 G} \int_{r_0}^{r_{\rm max}} \td r r^2 b \left[\frac{2}{r} + \frac{(b^2)'}{2 b^2} \right] \log \left\{ \frac{ 2 \alpha L_{ct} b }{r} \left[\frac{2}{r} + \frac{(b^2)'}{2 b^2} \right] \right\} \, .
\end{align}
This expression can be massaged into a more useful form via an application of integration by parts. Let us write the argument of the logarithm above as $\alpha L_{ct} F(r)$. Then we have:
\begin{align} 
I^F_{ct} &= \frac{\pi}{2 G} \int_{r_0}^{r_{\rm max}} \td r r^2 b \left[\frac{2}{r} + \frac{(b^2)'}{2 b^2} \right] \log \left( \alpha L_{ct} F(r) \right)\, ,
\\
&= \frac{\pi}{2 G} \left[r^2 b \log \left( \alpha L_{ct} F(r) \right) \right]_{r_0}^{\rmax} - \frac{\pi}{2 G} \int_{r_0}^{\rmax} \td r \frac{r^2 b F'(r)}{F(r)} \, .
\end{align}
Note in particular that the only dependence on the parametrization and the counterterm length scale $L_{ct}$ is in the terms evaluated at $r_0$ and $\rmax$, while the remaining integral is now independent of both of these quantities.\footnote{These manipulations are not specific to the spacetime considered here since in general we have $\sqrt{\gamma} \Theta = \td \sqrt{\gamma}/ \td\lambda$.} Furthermore, we can conclude that the term evaluated at $r_0$ actually vanishes --- the argument of the logarithm approaches a constant value, while $b(r) \to 0$ as $\sqrt{r-r_0}$. Therefore, only the contribution at $\rmax$ must be treated.

Let us note that the contribution due to the past null boundary can be computed in a completely analogous fashion. In that case the relevant normal one-form is
\be 
(k_P)_\mu \td x^\mu = \beta \left[-\td t + \td r^* \right] \, ,
\ee
where $\beta$ is a arbitrary (positive) constant associated with the parameterization. The computations go through exactly as before: the null boundary term vanishes, while for the counterterm we find that the only difference in the final answer is the substitution $\alpha \to \beta$. Noting this, we obtain
\be 
I^F_{ct}  + I^P_{ct} = \frac{\pi}{ 2 G} \left[r^2 b \log \left( \alpha \beta L_{ct}^2 F(r)^2 \right) \right]^{\rmax} - \frac{\pi}{ G} \int_{r_0}^{\rmax} \td r \frac{r^2 b F'(r)}{F(r)} \, ,
\ee
where we have made use of the fact that the first term vanishes when evaluated at $r_0$.

\subsubsection*{Joint contributions}

The WDW patch has two joint contributions arising where the cutoff surface $r = \rmax$ meets with the past and future null boundaries of the patch.\footnote{We are putting aside the joint terms arising due to the regulation of the caustics of the WDW patch --- these do not contribute, as discussed in appendix~\ref{furtherDeets}.} At these joints we must add the following contribution to the action:
\be 
I_{\rm jnt} = \frac{1}{8 \pi G} \int \td^3 x \sqrt{\sigma} a
\ee
where $\sigma$ is the determinant of the induced metric at the joint and for these particular joints $a$ is given by
\be 
a = \epsilon \log | \mathbf{k}_i \cdot \mathbf{n} |
\ee 
where the subscript $i$ stands for $F$ or $P$, depending on which null boundary of the WDW patch we are considering. To determine the factor $\epsilon$, we must first introduce an auxillary  timelike vector $\hat{t}^\mu \partial_\mu$ defined so that it lies within the tangent space of the boundary surface at $\rmax$ and is directed outward \textit{as a vector}. The factor $\epsilon$ is then obtained as
\be 
\epsilon = - {\rm sign}(\mathbf{k}_i \cdot \mathbf{n} ) {\rm sign}(\mathbf{k}_i \cdot \hat{\mathbf{t}} ) \, .
\ee

The outward-pointing normal one-form for the boundary at $\rmax$ is 
\be 
n_\mu \td x^\mu = \frac{1}{\sqrt{W}} \td r \, .
\ee
Consider now the joint occurring where the cutoff surface $\rmax$ intersects $F$. Here we introduce the auxilliary timelike normal
\be 
\hat{t}_\mu \td x^\mu = - \sqrt{\frac{r^2 W}{4 b^2}} \td t  \, ,
\ee
where the sign is chosen so that \textit{as a vector} it is outward pointing. Computing the relevant dot products we find that $\epsilon = -1$, and so
\be 
a^F_{r_{\rm max}} = - \frac{1}{2} \log \frac{4 \alpha^2 b^2 }{r^2 W} \, .
\ee

Running through the same analysis for $P$ we arrive at the same result up to the substitution $\alpha \to \beta$. Putting it all together, and noting that
\be 
\sqrt{\sigma} = \frac{r^2 b(r) \sin\theta}{4} \, , 
\ee
we have
\be 
I_{\rm jnt}^F + I_{\rm jnt}^P = - \frac{\pi r^2 b}{2 G} \log \frac{4 \alpha\beta b^2 }{r^2 W}  \, ,
\ee
where it is to be understood that this quantity is evaluated at $r = \rmax$. 

\subsubsection*{Bulk action}

Next we must evaluate the contribution from the bulk action on the WDW patch. After a short computation --- see appendix~\ref{furtherDeets} --- we find that the bulk action takes the form
\begin{align} 
I_{\rm bulk} &= - \frac{1}{16 \pi G} \int \td^5 x \sqrt{-g} \left[\frac{8}{\ell^2} + \frac{4 r_0^8}{j^2 r^6} \left[\frac{2}{r^2} + \frac{1}{\ell^2} - \frac{1}{j^2} \right] \right] \, ,
\\
&= - \frac{\pi}{4 G} \int_{r_0}^{\rmax} \td r r^3 \left[\frac{8}{\ell^2} + \frac{4 r_0^8}{j^2 r^6} \left(\frac{2}{r^2} + \frac{1}{\ell^2} - \frac{1}{j^2} \right) \right] \left(\frac{\pi \ell}{2} - r^*(r) \right) \, ,
\end{align}
where we have made use of our choice $v_\infty = - u_\infty = \pi \ell/2$.

Since we have to construct the tortoise coordinate numerically, this form of the bulk action is not so useful.  Fortunately, after integrating by parts we can put the bulk action into a more convenient form: 
\begin{align}
I_{\rm bulk} &= -\frac{\pi}{4 G} \int_{r_0}^{\rmax} \td r \frac{\td{\cal I}(r)}{\td r} \left(\frac{\pi \ell}{2} - r^*(r) \right)  \, ,
\\
&= -\frac{\pi}{4 G}  \left[{\cal I}(r) \left(\frac{\pi \ell}{2} - r^*(r) \right) \right]_{r_0}^{\rmax}-\frac{\pi}{4 G}  \int_{r_0}^{\rmax} \td r \frac{2{\cal I}(r)  b}{r W}  \, , 
\end{align} 
where 
\be\label{calI} 
{\cal I}(r) = \frac{2 r^4}{\ell^2} - \frac{2 r_0^8}{j^2 r^2} \left[\frac{1}{r^2} + \frac{1}{\ell^2} - \frac{1}{j^2} \right] \, .
\ee
This form of the expression is much more appropriate for numerical evaluation, since the tortoise coordinate does not explicitly appear inside an integral. Moreover, using the large-$r$ expansion of the tortoise coordinate
\be\label{bigTort} 
r^*(r) = \frac{\pi \ell}{2} - \frac{\ell^2}{r}  + \frac{\ell^4}{3 r^3} + \frac{j^2 \ell^2 p + 2 \ell^4(q-p) - \ell^6 }{5 r^5} + \cdots
\ee
we can show that the term ${\cal I}(r)(v_\infty - r^*(r))$ cancels with the analogous one arising from the global AdS solution in the limit $\rmax \to \infty$. Additionally, the contribution at $r = 0$ from this term for pure AdS vanishes.  And therefore this term contributes only at $r_0$.


\subsubsection{Complexity of formation}

Let us now put together the results we obtained above. Adding together the joint and counter terms and expanding the appropriate parts for large $\rmax$ we have
\be 
I^{F + P}_{\rm jnt} + I^{F + P}_{\rm ct} =  - \frac{\pi}{ G} \int_{r_0}^{\rmax}  \frac{r^2 b F'(r) \td r}{F(r)} + \frac{\pi}{4 G} \left[\rmax^3 \log \frac{9 L_{ct}^2}{\ell^2} + \ell^2 \rmax + {\cal O}\left(\rmax^{-1}\right) \right] \, ,
\ee
where
\be 
F(r) = \frac{ 2 }{r} \left[b' + \frac{2 b}{r}  \right] \, .
\ee
The result is manifestly independent of the parameterization of the null generators, depending only on the counterterm scale $L_{ct}$. While this conclusion is necessary, it nonetheless provides a consistency check of our computations.  The analogous sum of terms for the AdS vacuum corresponds to the above result upon setting\footnote{We assume here that the value of $L_{ct}$ is the same for both the soliton spacetime and the AdS vacuum.} 
\be 
F_{\rm AdS}(r) = \frac{3}{r} \, , \quad  b_{\rm AdS}(r) = \frac{r}{2} \, ,
\ee 
and $r_0 = 0$. It is therefore clear that the dependence on the counterterm length scale drops out when the subtraction is performed. We end up with the  following for the difference of the counterterm and joint contributions:\footnote{Let us mention that the joint contributions independently cancel when the subtraction is performed. However, the advantage of working with the sum of the joint and counterterm contributions is in the fact that it makes manifest the parameterization independence of the action for both spacetimes. Ultimately, it is the counterterm that yields a finite contribution to the complexity of formation.}
\be 
\Delta  \left(I_{\rm ct} + I_{\rm jnt} \right) = -\frac{\pi r_0^3}{6 G} - \frac{\pi}{ G} \int_{r_0}^{\infty}  \left[\frac{r^2 b F'(r)}{F(r)} + \frac{r^2}{2} \right] \td r \, .
\ee
With the subtraction of the AdS contribution, the remaining integral is convergent and so we have extended the integration to $\rmax = \infty$. 

We can perform also the background subtraction for the bulk contribution. In this case we have
\begin{align} 
\Delta I_{\rm bulk} =& \frac{\pi}{4 G} \left[{\cal I}(r_0) \left(\frac{\pi \ell}{2} - r^*(r_0) \right) \right] + \frac{\pi\left[r_0^3 - 3 \ell^2 r_0 + 3 \ell^3 \arctan (r_0/\ell) \right]}{6 G} 
\nn\\
&- \frac{\pi}{2 G} \int_{r_0}^\infty \left[\frac{ {\cal I}(r) b(r)}{r W(r)} - \frac{ r^4}{\ell^2 + r^2} \right] \td r \, ,
\end{align}
here, as before,
\be 
{\cal I}(r) = \frac{2 r^4}{\ell^2} - \frac{2 r_0^8}{j^2 r^2} \left[\frac{1}{r^2} + \frac{1}{\ell^2} - \frac{1}{j^2} \right] \, .
\ee

\begin{figure}[tp]
\centering
\includegraphics[width=0.6\textwidth]{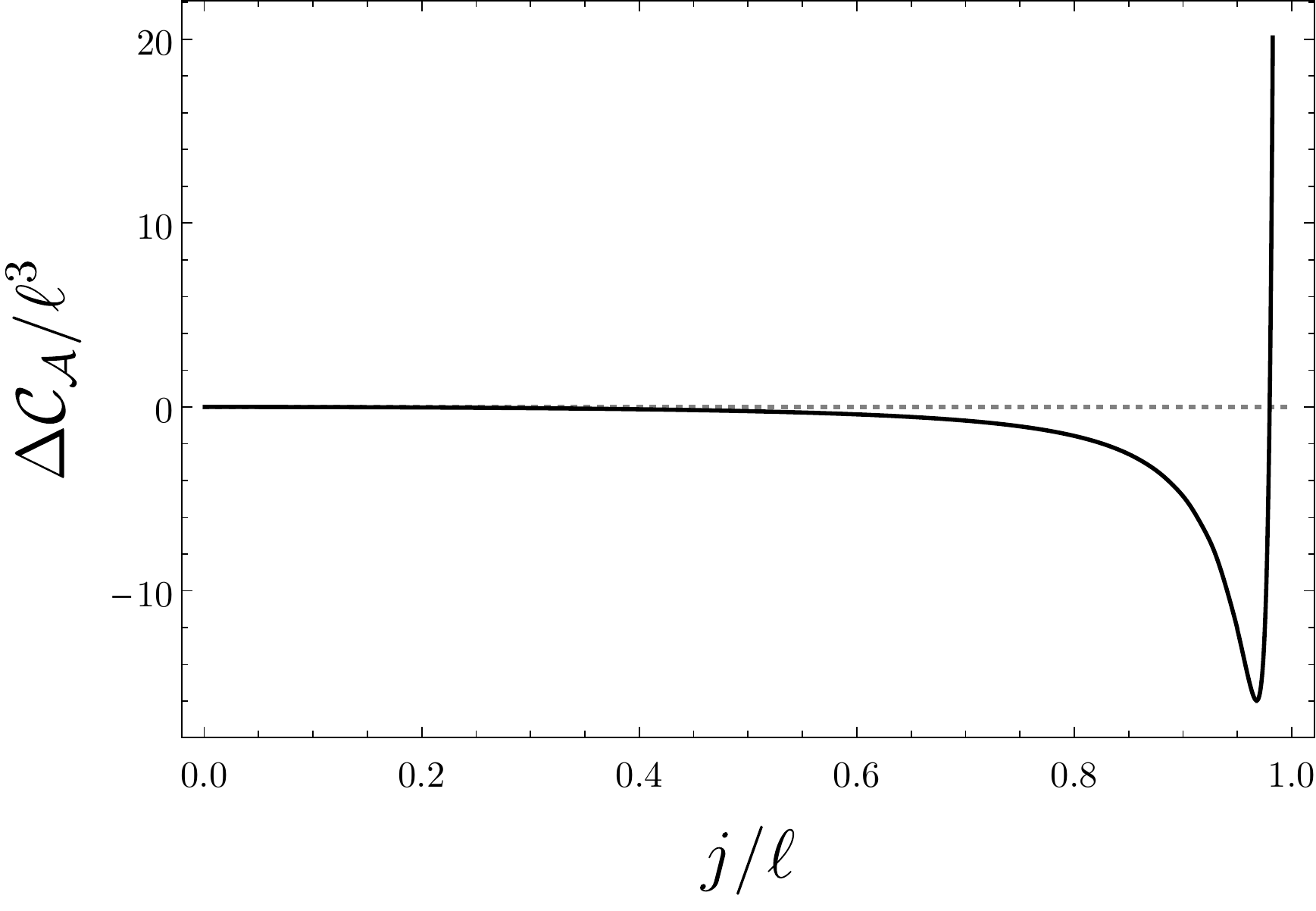}
\caption{Complexity of formation for the CA proposal. In sharp contrast to the CV proposal, the complexity of formation is neither strictly positive nor monotonic.}
\label{CAform}
\end{figure}

The complexity of formation for the soliton is then expressed as
\be 
\Delta {\cal C} = \frac{\Delta I_{\rm WDW}}{\pi} \,
\ee
which we can now evaluate numerically. Doing so produces the curve shown in figure~\ref{CAform}. The result is vastly different from what was observed for the CV proposal. First, note that the complexity of formation is not strictly positive, but is in fact negative for a large portion of the parameter space. Furthermore, the complexity of formation is not monotone, instead exhibiting a global minimum. Mathematically, this behaviour is due to the behaviour of the counterterm. On its own, the bulk action is a strictly negative, monotonically decreasing function. Meanwhile the counterterm is a manifestly positive, increasing function. When the solitons become sufficiently large, the positive growth of the counterterm overwhelms the negative growth of the bulk term, resulting in the minimum observed. The fact that the complexity of formation crosses through zero a second time indicates that there is a particular soliton configuration that is of equal complexity to the vacuum.

It is worth noting that the complexity of formation is sensitive to features due to angular momentum, such as ergoregions, though this is not the dominant effect. In terms of the functions appearing in the metric, the integrand of the bulk action can be written as --- see appendix~\ref{furtherDeets} ---   
\be 
-\frac{1}{r^3}\frac{\td \mathcal{I}(r)}{\td r} = -\frac{8}{\ell^2} - \frac{4 q^2 j^2}{r^8} + \frac{16 q^2 b(r)^2(1+ j f(r))}{r^8} + \frac{4 q^2 j^2 g_{tt}}{r^8} \, .
\ee
The impact of the last three terms is magnified due to the prefactor $q$ that appears --- recall that regularity fixes $q = -r_0^4/j^2$. The limit $j \to \ell$ corresponds to $r_0 \to \infty$ and as a result these terms become increasingly important for large solitons.
Since for large enough $j$ the spacetime possesses ergoregions $g_{tt}$ is not strictly negative, which has the effect of increasing the numerical value of the above term. This in turn decreases the value of the bulk contribution to the complexity of formation. In other words --- at least in this case --- ergoregions decrease the complexity of formation, albeit only slightly.

Unfortunately, it is more difficult to perform a perturbative analysis of the integrals here. The reason has to do with the fact that various terms appearing in the integral for the bulk and the counterterm have dependence $(r-r_0)^{-1/2}$. As a result, these expressions are integrable but they are not differentable at $r=r_0$. Therefore we cannot integrate the series of the integrand for small $r_0$ as we did for the CV proposal. Nonetheless, we can extract useful asymptotics numerically. Like was the case for the CV proposal, we find that for large solitons the behaviour proportional to $V^{3/4}$ with a dimensionless prefactor:
\be 
\Delta\mathcal{C}_\mathcal{A} \approx 0.1617 V^{3/4} \, .
\ee
It is also insightful to understand the behaviour of the bulk action and the counterterm individually. For these we find that
\be 
\Delta I_{\rm bulk} \approx -0.7948 V^{3/4} \, , \quad \Delta I_{\rm ct} \approx 1.3027 V^{3/4}
\ee
in the limit of large solitons.

\section{Discussion}
\label{discuss}

Although much of the interest in gravitational solitons arose as a consequence of the proposal that they are microstate geometries for black holes, it is clear that they should be considered as important classical solutions in their own right, characterized by nontrivial spacetime topology.  Indeed in the asymptotically flat case they satisfy a Smarr-type relation as well as a `first law' variation formula where the independent extensive variables are now the various fluxes that support the soliton, rather than the angular momenta and horizon area as in the familiar case for black holes.   The purpose of this work was to address the situation in the asymptotically globally AdS$_5$ setting and compare the behaviour of these solitons with that of black holes.  To do so, we have focussed attention on family of cohomogeneity-one solitons, which helps simplify the analysis.   We have found some important differences as we summarize below. 

Firstly, we have shown that the solitons satisfy a mass formula and first law mass variation formula. In both cases, a thermodynamic volume plays an important role in the construction. In contrast to more commonly studied examples involving black holes, here the thermodynamic volume arises due to the nontrivial topology of the spacetime. That is, it is not possible to find a Killing potential that is both regular through the entire spacetime and regularizes the Komar integral to yield the correct AMD mass at infinity. Choosing to satisfy the latter condition in turn leads to the association of a nontrivial thermodynamic volume to the bubble. 

Along these lines there are a few natural directions to proceed. The mathematical considerations here could be applied to other soliton spacetimes, such as the AdS soliton~\cite{Horowitz:1998ha}, the Eguchi-Hanson solitons~\cite{Clarkson:2005qx, Mbarek:2016mep}, or the solitons considered in~\cite{Compere:2009iy} which can be thought of as the generalization of the solitons we have studied to the case of nonequal angular momenta. More ambitiously, it should be possible to apply the Hamiltonian perturbation theory arguments of~\cite{Kastor:2009wy} to generalize the results of~\cite{Kunduri:2017htl} to the asymptotically AdS setting.  Such a construction would then be valid for general soliton-black hole configurations. In this case, it seems natural to expect that the total thermodynamic volume of the spacetime would consist of a sum of topological contributions from solitons in addition to the familiar contribution from the black hole horizon. It would be interesting to understand the implication of these terms for the conjectured `reverse' isoperimetric inequality of~\cite{Cvetic:2010jb}.

In addition, we also briefly considered the existence of ergoregions, evanescent ergosurfaces, and stable trapping of null geodesics.  There is indeed a critical value in the parameter space of solutions at which an evanescent ergosurface forms, and beyond which there is both an inner and outer ergosurface.  This suggests instabilities should occur, at least based on intuition from the asymptotically flat case. Interestingly, the BPS member in this class of solitons do not exhibit either of these properties, although they do allow for stable trapping of null geodesics. A thorough investigation of instabilities of these AdS$_5$ solitons, and the possible stationary endpoints, remains an open problem.

We have also considered these soliton metrics in light of the holographic ``complexity equals volume'' and ``complexity equals action'' conjectures. In each case the the complexity is independent of time --- gravitationally, this is due to the fact that the solitons do not have horizons or ``interiors'', while from the CFT perspective it is consistent with the idea that these solitons should correspond to energy eigenstates of the CFT hamiltonian. We have computed the complexity of formation relative to the AdS vacuum. In both cases we find that the complexity of formation for large solitons is directly proportional to a power of the thermodynamic volume $\Delta \mathcal{C}_{\mathcal{A}, \mathcal{V}} \propto V^{3/4}$. This particular dependence on $V$ agrees exactly with that for black holes, up to numerical prefactors. In the black hole case it is perhaps more natural to write the complexity of formation in terms of the entropy, which has a clear interpretation as a measure of the degrees of freedom  of the dual theory.\footnote{It is worth noting that the complexity of formation for large solitons could equally well be expressed as proportional to $\ell^\# M^{3/4}$. However, we emphasize that the mass alone does not properly account for the combination of $r_0$ and $\ell$ dependence that appears in the complexity of formation, while the thermodynamic volume does.} However, the soliton metrics have zero entropy, and therefore it is somewhat surprising that the same functional form holds when written in terms of the thermodynamic volume.  It seems that this connection may warrant further investigation. 

Unfortunately, the holographic interpretation of the thermodynamic volume is much less complete --- see~\cite{Johnson:2014yja, Karch:2015rpa, Johnson:2018amj} for remarks on this --- and, therefore, it is difficult to draw concrete conclusions regarding the implications of this fact for the dual CFT. However, let us note that there have been proposals advocating the importance of the thermodynamic volume in holographic complexity, e.g. ``complexity equals volume 2.0''~\cite{Couch:2016exn}. In that context it was argued that the thermodynamic volume provides a probe of the black hole interior.  One observation is the following. Suppose we were to regard these solitons as corresponding to some type of black hole microstates.\footnote{This possibility was suggested in~\cite{Ross:2005yj}.} Then note that for large solitons (in each proposal) $\Delta \mathcal{C} \sim N_c^\# (V/\ell^4)^{3/4}$  where $N_c$ is related to the central charge of the dual CFT. Thermofield double states corresponding to black holes with entropy $S$ would require $e^S$ microstates. Since the complexity of formation for black holes is proportional to their entropy, it is far more computationally efficient --- \textit{by a factor of $e^S$} --- to construct the thermofield double states directly rather than first constructing the energy eigenstates from which they are comprised.

Beyond these similarities for large solitons, the properties of the complexity of formation is drastically different for the two proposals. We found that the complexity of formation within the ``complexity equals volume'' proposal is a positive, monotonically increasing function of the mass (equivalently angular momentum). The monotonicity is  consistent with other works considering the complexity of formation for smooth geometries~\cite{Reynolds:2017jfs, Bombini:2019vuk}. However, note that in the case of the AdS soliton~\cite{Reynolds:2017jfs} $\Delta \mathcal{C}_\mathcal{V}$ was found to be monotonic but \textit{decreasing}. This difference is likely due to the fact that the AdS soliton has negative mass, while the solitons considered here have positive mass. Overall, the CV proposal does not appear to be sensitive to the nontrivial rotational properties of the spacetime. On the other hand, the action  exhibits a surprisingly nontrivial dependence on the angular momentum. For small solitons, $\Delta \mathcal{C}_{\mathcal{A}}$ is a monotonically decreasing function of the mass/angular momentum. However, for sufficiently large angular momentum the complexity of formation reaches a minimum before diverging to $+\infty$ as $r_0 \to \infty$: the action is not strictly positive nor monotonic. Comparing this result with that of the AdS soliton~\cite{Reynolds:2017jfs},
 the behaviour of small solitons is qualitatively similar but again in opposite fashion (for the AdS soliton $\Delta\mathcal{C}_\mathcal{A}$ is positive and increasing), but the behaviour of large solitons in markedly different. Mathematically, the reason for the nonmonotonicity arises from the interplay of the bulk and counterterm contributions to the action. On its own, the bulk term is a negative, monotonically decreasing function. However, the counterterm is manifestly positive and monotonically increasing. While for small angular momentum the bulk action wins, ultimately the positive growth of the counterterm overwhelms the negative growth of the bulk action. It is important to note that, assuming the ambiguous counterterm length scale $L_{\rm ct}$ is the same for both the soliton and the AdS vacuum, the conclusions just described are completely independent of the choice of this parameter. 

It is natural to wonder if with a suitable choice of additional counterterms the action and volume can be restored to equal footing, similar to the considerations of dyonic black holes in~\cite{Brown:2018bms, Goto:2018iay}. Obviously one possibility would be to simply neglect the counterterm contribution, which would restore monotonicity. However this seems unappealing due to the recent observations vouching for the importance of the counterterm for reproducing expected properties of the complexity, e.g.,~\cite{Chapman:2018dem, Chapman:2018lsv}. It would therefore seem unnatural to have a prescription that includes the counterterm in one setting while neglecting it in another.  Regarding other options for counterterms, while we cannot rule out this possibility, we have not succeeded in producing such an effect with simple choices. One option is the inclusion of a GHY-type term on the surface of the bubble. While non-vanishing, a term of this kind is not really warranted in the present situation due to the smoothness of the geometry at the location of the bubble.\footnote{Note that in contrast to the case of a black hole singularity, the surface $r = r_0$ is not a boundary of the WDW patch here. Including a surface term there is possible, but would be morally similar to including a surface term at the black hole horizon in the computation of the Euclidean action.} In any case, such a term does not restore monotonicity to the action. We have also explored --- see appendix~\ref{furtherDeets} --- the implications of the electromagnetic terms specifically. They make a negative contribution to the complexity of formation. Positivity of the complexity of formation could be restored by  subtracting a large, negative multiple of these electromagnetic terms, however such a procedure would seem to lack physical motivation. It is also possible that the complexity of formation is legitimately nonmonotonic and the action is capturing the correct behaviour. In this case, our results would provide an example where one complexity conjecture would be favoured over the other --- see~\cite{Reynolds:2017jfs, Chapman:2018bqj, Cooper:2018cmb, Ross:2019rtu} for other results on this subject --- and another example of a situation where the counterterm plays an essential role. Unfortunately, without a better understanding of the CFT states to which these bulk solutions correspond we cannot determine which conjecture behaves more favorably.


We have also noted a weak but nontrivial dependence of the action on the existence of ergoregions in the spacetime. In the present analysis, ergoregions tend to decrease the complexity of formation. To the best of our knowledge this is the first example of a genuine feature of rotating spacetimes having an impact within the holographic complexity conjectures.  Here this is possible due to the presence of the Maxwell field --- we do not expect ergoregions to affect the numerical value of the action in vacuum spacetimes. It would be interesting to further explore this feature for charged and rotating black holes, and especially to determine if it is natural from a CFT perspective.

Before closing, it is worth mentioning the precise reason why the action computation was far simpler for these solitons than one would expect for a spacetime with general rotation. The reason has to do with the fact that, for equal angular momenta, the dependence on the polar angle in the metric drops out rendering the solution cohomogeneity-one. Presumably a similar simplification occurs for equal-spinning higher-dimensional rotating black holes and could perhaps be leveraged to explore the holographic complexity conjectures in that context.

\section*{Acknowledgements} 

\noindent We thank Pablo Bueno, Pablo Cano, Richard Epp, Graham Cox, David Kubiznak, James Lucietti, Robb Mann, Hugo Marrochio, and Fil Simovic for useful comments and discussions. We thank Hugo Marrochio for comments on an earlier version of this manuscript. RAH is grateful to the University of Waterloo and Perimeter Institute for Theoretical Physics for hospitality during the completion of part of this work. HKK is supported by an NSERC Discovery Grant. The work of RAH is supported by NSERC through the Banting postdoctoral fellowship program. This research was supported in part by Perimeter Institute for Theoretical Physics. Research at Perimeter Institute is supported by the Government of Canada and by the Province of Ontario. 

\appendix

\section{The $r_0 \to 0$ limit of the metric}
\label{smallSol}

Here we will collect a few details regarding the behaviour of the metric in the limit $r_0 \to 0$. To take this limit, we first introduce a parameter 
\be 
y \equiv \frac{r_0}{\ell} \, , \quad y \ll 1 \, .
\ee
We demand that the regularity of the metric is respected as $y \to 0$. This constrains the functional form of $j$ according to eq.~\eqref{reg}. Since the constraint is a cubic it can be solved exactly, however a series representation for small $y$ is more useful. A direct expansion of the closed form result reveals that
\be 
j^2 = \ell^2 \left[j_0^2 y^2 - \frac{j_0^4(1+2j_0^2)}{3} y^4 + \frac{j_0^8(1+2j_0^2)}{3} y^6 - \frac{573 j_0^4 + 717 j_0^2 + 194}{729} y^8 + \cdots  \right]
\ee
where $j_0$ is one of the two real solution to the equation
\be 
3 j_0^6 - 3 j_0^2 - 1 = 0 \, , \quad  \Rightarrow \quad  j_0 \approx \pm 1.06638 \, .
\ee
Note that we have used this result to simplify the expansion for $j^2$.

\begin{figure}[!ht]
\centering
\includegraphics[width=0.65\textwidth]{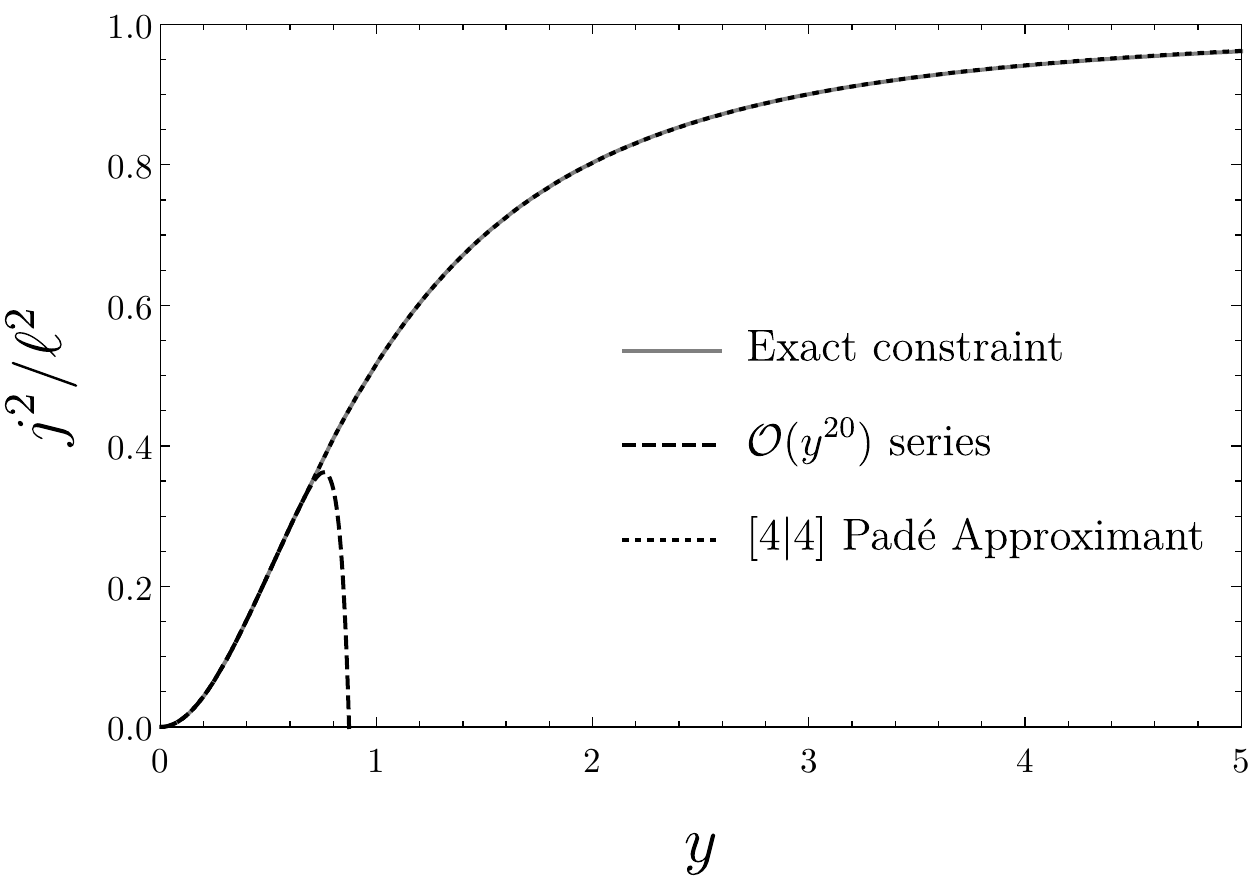}
\caption{A plot comparing the exact constraint~\eqref{reg} (solid, gray curve) with approximations constructed for small $y$: The dashed black curve corresponds to a power series representation including terms up to ${\cal O}(y^{20})$, while the dotted black curve corresponds to a $[4|4]$ Pad{\'e} approximant. The Pad{\'e} approximant is virtually indistinguishable from the original curve.}
\label{jExpandPlot}
\end{figure}

Using the ratio test on the explicit solution it can be confirmed that the radius of convergence for the power series is $3/4$. However, we note that even a low order Pad\'e approximant displays much better convergence properties, as we illustrate in figure~\ref{jExpandPlot}. In this example we use a $[4|4]$ Pad\'e approximant, which has the explicit form
\be 
\frac{j^2}{\ell^2} = \frac{3 y^2 \left[-9 + 54 j_0^4 + 2(-2 + j_0^2 + 15 j_0^4) y^2 \right]}{81(1 + 2 j_0^2 - j_0^4) + 3 ( 23 + 60 j_0^2 + 6 j_0^4) y^2 + (12 45 j_0^2 + 37 j_0^4) y^4} \, .
\ee

To understand the limiting behaviour of the metric, we explicitly expand the metric functions to leading order in $y$. We have
\begin{align}
b^2(r) &= \frac{r^2}{4} - \frac{(-1 + j_0^2)\ell^4}{4 j_0^2 r^2} y^4 + {\cal O}(y^6) \, ,
\\
W(r) &= 1 + \frac{r^2}{\ell^2} - \frac{(1+j_0^2) \ell^2}{j_0^4 r^2} y^2 + {\cal O}(y^4) \, ,
\\
f(r) &= -\frac{2 \ell^3}{j_0^3 r^4} y^3 + {\cal O}(y^5) \, .
\end{align}
These expansions of the metric functions, along with the fact that we have demanded that the regularity holds, makes clear that as $r_0 \to 0$ the metric limits smoothly to AdS$_5$. Computing the curvature (using Maple) and working in the $y \to 0$ limit, we find that $R_{ab}^{cd} = -2 \ell^{-2} \delta_{[a}^c \delta_{b]}^d + {\cal O}(y^2)$, confirming the above conclusion.

\section{Fefferman-Graham form \& holographic stress tensor}\label{FGstress}

In the process of calculating the complexity of formation it was necessary to subtract the divergent contribution due to pure AdS. Here, we collect some results concerning the asymptotic expansion of the metric, which allows for a more careful matching of the geometries in the large-$r$ regime. As a by-product, we are able to obtain an expression for the holographic stress tensor, which provides an independent confirmation of the conserved charges advocated in the main text.

Let us define a new coordinate $\rho$ in the following way:
\be 
\frac{\ell^2}{\rho^2} \td \rho^2 \equiv \frac{\td r^2}{W} \, .
\ee
With this definition, we can write $r$ as a function of $\rho$. The result, working only to the order at which the soliton metric begins to differ from global AdS is
\be 
r = \rho - \frac{\ell^2}{4 \rho} + \frac{\ell^2(p-q) - j^2 p}{4 \rho^3} + {\cal O}( \rho^{-5}) \, .
\ee
Substituting this new coordinate into the metric, we find that it takes the Fefferman-Graham form
\be 
\td s^2 = \frac{\ell^2}{\rho^2} \td\rho^2 + \gamma_{\mu\nu} \td x^\mu \td x^\nu
\ee
where $\gamma_{\mu\nu}$ can be expressed in the following way:
\be 
\gamma_{\mu\nu} = \frac{\rho^2}{\ell^2} \left[\gamma_{\mu\nu}^{(0)} + \frac{\ell^2}{\rho^2} \gamma_{\mu \nu}^{(2)} + \frac{\ell^4}{\rho^4} \gamma_{\mu \nu}^{(4)} + \cdots \right] \, .
\ee
Here, the AdS boundary is located at $\rho = \infty$. However, note that the transformation $z\equiv \ell^2/\rho$ leaves the form of the metric invariant, while mapping the boundary to $z = 0$. Next, by performing the explicit expansions we obtain the following for the $\gamma_{\mu\nu}^{(i)}$:
\begin{align}
\gamma^{(0)}_{\mu\nu}\td x^\mu \td x^\nu &=  - \td t^2 + \frac{\ell^2}{4} \left[\sigma_1^2 + \sigma_2^2 + \sigma_3^2 \right] \, ,
\\
\gamma^{(2)}_{\mu\nu}\td x^\mu \td x^\nu &= - \frac{1}{2}\td t^2 - \frac{\ell^2}{8}   \left[\sigma_1^2 + \sigma_2^2 + \sigma_3^2 \right] \, ,
\\
\gamma^{(4)}_{\mu\nu}\td x^\mu \td x^\nu &= \frac{24 \ell^2 (p-q) + 8 j^2 p - \ell^4}{16 \ell^4} \td t^2 + \left(\frac{\ell^4 - 8 j^2 p + 8 \ell^2(p-q)}{64 \ell^2} \right) \left[\sigma_1^2 + \sigma_2^2 \right] 
\nonumber\\
&+ \left(\frac{\ell^4 + 24 j^2 p + 8 \ell^2 (p-q)}{64 \ell^4} \right)\sigma_3^2 + \frac{j(q-2p)}{\ell^2} \sigma_3\td t \, .
\end{align}

Having this expansion, we can do a couple of useful things. First, we are able to confirm that setting a cutoff surface at $r = r_{\rm max}$ will yield no discrepancy when performing a subtraction between the soliton spacetime and global AdS. The reason is simply because of the following. If we place a UV cutoff at $\rho = \ell^2/\delta$ then 
\be 
r_{\rm max} - r_{\rm max}^{\rm AdS} = \frac{\ell^2(p-q) - j^2 p}{4 \ell^6} \delta^3 + {\cal O}(\delta^5) \, .
\ee
This difference approaches zero in the limit $\delta \to 0$, and therefore there will be no finite contribution arising from the matching of the geometries near the boundary.

The second useful thing the above expansion allows us to do is obtain an independent confirmation of the thermodynamic mass presented in the main text via a computation of the holographic stress tensor~\cite{Balasubramanian:1999re} (see also~\cite{Brown:1992br}). To this end, let us recall that the gravitational part of the action --- including suitable boundary and counterterms to render the on-shell action finite --- is given by~\cite{Emparan:1999pm}\footnote{Note that our expression differs from that in \cite{Emparan:1999pm} by an overall sign, this is simply because we are working here with the Lorentzian action. }
\begin{align}
I_{\rm grav} =& \frac{1}{16 \pi G} \int_{\mathcal M} \td^5x \sqrt{-g} \left[R - 2 \Lambda \right]
 + \frac{1}{8 \pi G} \int_{\partial \mathcal{M}} \td^4 x \sqrt{-\gamma} \left[K - \frac{3}{\ell} - \frac{\ell}{4} {\cal R} \right] \, ,
\end{align}
where here we have  the extrinsic curvature $K = \gamma^{\mu\nu} K_{\mu\nu}$ with $K_{\mu \nu} \equiv \gamma_{\mu}^\alpha \nabla_\alpha n_\nu$, while ${\cal R}$ is the Ricci scalar computed for the boundary metric $\gamma_{\mu\nu}$. Note that our conventions for the extrinsic curvature here are the same as those in the main text: we demand that that the \textit{one-form} normal to the surface is \textit{outward pointing}.

The holographic stress tensor is defined according to the usual formula
\be 
T_{\mu\nu} \equiv -\frac{2}{\sqrt{-\gamma}} \frac{\delta I_{\rm grav}}{\delta \gamma^{\mu\nu}} \, .
\ee
Performing the variation of the action, we obtain the following result
\be 
8 \pi  G T_{\mu\nu} = -K_{\mu\nu} + K \gamma_{\mu\nu} - \frac{3}{\ell}\gamma_{\mu\nu} + \frac{\ell}{2} {\cal G}_{\mu\nu} \, ,
\ee
where $K_{\mu\nu} = (\rho/\ell) \partial_\rho \gamma_{\mu\nu}$ is the extrinsic curvature of the boundary and ${\cal G}_{\mu\nu}$ is the Einstein tensor for the metric $\gamma_{\mu\nu}$. Note that the signs appearing in the above expression differ from those in~\cite{Balasubramanian:1999re} simply because we are using the more common conventions for the definition of the curvature tensor and extrinsic curvature.

The general form of the stress tensor is not particularly enlightening, and so we do not present it here --- we note, though, that does not take the form of a simple perfect fluid. Nonetheless, from the holographic stress tensor we can compute the conserved mass in the following way:
\be 
M = \int_{\Sigma} \td^3 x \sqrt{\sigma} u^\mu \xi^\nu T_{\mu \nu} \, ,
\ee
where $\Sigma$ is a constant $t$ hypersurface in the boundary metric, $u^\mu$ is the future-pointing timelike unit normal vector to $\Sigma$, $\sigma$ is the determinant of the induced metric on $\Sigma$, and $\xi^\mu$ is the time-like Killing vector. In our case, we have $u^\mu = (\rho/\ell) \delta_t^\mu$ and $\sqrt{\sigma} = (\rho^3 \sin\theta)/8$,  we take $\xi^\alpha =  \delta^\alpha_t$, and  the relevant component of the stress tensor reads
\be 
T_{tt} = \frac{3 \ell^4 + 8 j^2 p + 24 \ell^2(p-q)}{64 \pi G \ell^3 \rho^2} + {\cal O}(\rho^{-3}) \, .
\ee
Performing the computation and taking the limit $\rho \to \infty$ we obtain
\be 
M = \frac{\pi\left[3(p-q)\ell^2 + j^2 p \right]}{4 \ell^2 G} + \frac{3 \pi \ell^2}{32 G} \, .
\ee
The first term appearing in the above result is precisely that which we obtained before using the Ashtekar-Magnon procedure. The additional contribution seen here  corresponds to the Casimir energy of AdS$_5$. 

We can also compute the conserved angular momenta from the holographic stress tensor. In this case, the relevant conserved charge is
\be 
J_i = \int_{\Sigma} \td^3 x \sqrt{\sigma} u^\mu \eta_{(i)}^\nu  T_{\mu \nu}
\ee
where $\eta_{(i)}^\mu = \sigma^{\mu\nu} (\eta_{(i)})_\nu$ with $(\eta_{(i)})_\nu = \delta^i_\mu$ is an angular Killing vector. Performing the computations, we find that
\be 
J_\psi = \frac{j \pi (2p-q)}{4 G} \, , \quad J_\phi = 0 \, ,
\ee
in agreement with the expressions~\eqref{angMom}.

\section{Further details for action computation}
\label{furtherDeets}

\subsection{No contributions from caustics}

Here we show that the caustics at the future and past tips of the WDW patch do not contribute to the action. 

Our computation is analogous to the one for the AdS vacuum in~\cite{Chapman:2016hwi}.  Specifically, we eliminate the caustic by introducing spacelike hypersurfaces at $t = v_\infty - r^*(r_0 + \epsilon)$ and $t = u_\infty + r^*(r_0 + \epsilon)$. We then evaluate the GHY and corner contributions in the limit $\epsilon \to 0$. 

Consider the future tip of the WDW patch. In this case the relevant outward pointing normal 1-form is 
\be 
t_\mu \td x^\mu = \sqrt{\frac{r^2 W}{4 b^2}} \td t \, .
\ee
It is simple to verify that the trace of the extrinsic curvature of this surface vanishes (essentially due to the fact that the spacetime is stationary), and therefore there is no boundary contribution there.

To compute the joint contribution where the cutoff surface intersects the future light-like boundary of the WDW patch, we introduce an auxillary spacelike outward pointing \textit{vector}:
\be  
\hat{n}^\mu \partial_\mu = \sqrt{W} \partial_r \, .
\ee
Taking the relevant dot products, we find that $\epsilon = +1$ and therefore we have
\be 
a = \frac{1}{2} \log \frac{4 \alpha^2 b^2}{r^2 W} \, , 
\ee
and since
\be 
\sqrt{\sigma} = \frac{r^2 b \sin\theta}{4} \, ,
\ee
we have
\be 
I_{\rm jnt} =  \lim_{r\to r_0} \frac{\pi^2 r^2 b(r) }{4}  \log \frac{4 \alpha^2 b^2}{r^2 W} = 0 \, . 
\ee
A completely analogous analysis holds for the past tip of the WDW patch. Therefore we conclude that, just as in the vacuum case, the caustics of the WDW patch make no contribution to the action.

\subsection{Computation of bulk action}

Using the field equations, the on-shell action is 
\begin{equation}
I_{\rm bulk} = \frac{1}{16\pi G} \int_{\mathcal{M}} \left[ -\frac{2}{3} |F|^2 - \frac{8}{\ell^2}\right] \td \text{Vol}(g) - \frac{8}{3\sqrt{3}} F \wedge F \wedge A
\end{equation} where we used the identity $2 F \wedge \star F = |F|^2 \td \text{Vol}(g)$.  


To compute $|F|^2$ we note that $\star F$:
\begin{equation}
\star F = \frac{\sqrt{3} q}{2} \left[ -\frac{j \sqrt{W}}{r^3 b(r)} e^0 \wedge e^3 \wedge e^4 + \frac{2 b}{r^4} (2 + j f(r)) e^2 \wedge e^3 \wedge e^4 - \frac{2 j}{r^4} e^0 \wedge e^1 \wedge e^2 \right ] 
\end{equation}  where we use the orthonormal  basis
\begin{align}
e^0  &= \frac{r \sqrt{W(r)}}{2 b(r)} \td t ,  \qquad e^1 = \frac{ \td r}{\sqrt{W(r)}}, \quad e^2 = b(r) (\td \psi + \cos\theta \td \phi + f(r) \td t), \\ e^3 &= \frac{r}{2} \td \theta, \quad e^4 = \frac{r}{2} \sin \theta \td \phi \, .
\end{align}

 Then $2F \wedge \star F = |F|^2 \td \text{Vol}(g)$ gives
\begin{equation}
|F|^2 = \frac{3q^2}{2} \left[ \frac{j^2 W}{r^6 b(r)^2} + \frac{4j^2}{r^8} - \frac{4 b(r)^2}{r^8}(2 + j f(r))^2 \right] \, .
\end{equation}  The other term appearing in the action is $F \wedge F \wedge A$.  We find
\begin{align}
F \wedge F & = \frac{3 q^2 j}{r^7}\left[ \frac{ j \sqrt{W}}{b(r)} e^1 \wedge e^2 \wedge e^3 \wedge e^4 + \frac{2 b(r)}{r} (2 + j f(r)) e^0 \wedge e^1 \wedge e^3 \wedge e^4 \right] 
\end{align} and in the gauge where $A \to 0$ at spatial infinity
\begin{equation}
A = \frac{\sqrt{3} q j }{4r^2 b(r)} e^2 - \frac{\sqrt{3} q b(r)}{r^3 \sqrt{W}}\left( 1 + \frac{jf(r)}{2}\right) e^0 \, .
\end{equation} We then find $F \wedge F \wedge A =0$ identically.

If we explicitly evaluate $|F|^2$ we find
\begin{equation}
|F|^2 = \frac{6 r_0^8}{j^2 r^6} \left[\frac{2}{r^2} + \frac{1}{ \ell^2} - \frac{1}{j^2}\right] \, .
\end{equation} The on shell action is then
\begin{equation}
I_{\rm bulk}  = \frac{1}{16\pi G} \int_{\mathcal{M}} \left( -\frac{4 r_0^8}{j^2 r^6} \left[\frac{2}{r^2} + \frac{1}{ \ell^2} - \frac{1}{j^2}\right] - \frac{8}{\ell^2}\right) \td \text{Vol}(g) \, .
\end{equation}

\subsection{Vanishing of GHY term}

With our regularization of the WDW patch, there is a GHY contribution on the timelike surface $r = \rmax$. Here we will show that this contribution is precisely canceled by the analogous contribution arising for the pure AdS vacuum.

The outward-pointing normal one-form at the $r = \rmax$ surface is given by
\be 
n_\mu \td x^\mu = \frac{1}{\sqrt{W}} \td r \, .
\ee
The extrinsic curvature of this surface is easily computed as
\be 
K = \nabla_\mu n^\mu  = \frac{1}{\sqrt{-g}} \left(\sqrt{-g}n^\mu \right)_{,\mu}  = \frac{\sqrt{W}}{2} \left[\frac{6}{r} + \frac{W'}{W} \right] \, .
\ee

The GHY contribution is obtained by integrating this expression over the boundary segment. In this case, the $\rmax$ surface meets the past null boundary of the WDW patch at $t = u_\infty + r^*(\rmax)$ and the future null boundary at $t= v_\infty - r^*(\rmax)$. Noting that the determinant of the induced metric on this surface is
\be 
h = - \frac{r^6 W(r) \sin^2\theta}{64}
\ee
we obtain
\be 
I_{\rm GHY}^{\rmax} = \frac{\pi \rmax^3}{4  G}  \left[W'(\rmax) + \frac{6 W(\rmax)}{\rmax}  \right]\left(\frac{\pi \ell}{2} - r^*(\rmax) \right) \, .
\ee

The analogous term for global AdS can be obtained from the result above upon setting
\begin{align} 
W(r) &= 1 + \frac{r^2}{\ell^2} \, ,
\end{align}
and replacing the tortoise coordinate with the corresponding one for AdS
\be 
r^*_{\rm AdS}(r) = \ell \arctan \frac{r}{\ell} \, .
\ee
Using this result, along with the asymptotic form of the tortoise coordinate~\eqref{bigTort}, we find that
\be 
I_{\rm GHY}^{\rmax} - \left(I_{\rm GHY}^{\rmax} \right)_{\rm AdS} = {\cal O} \left( \rmax^{-1} \right) \, .
\ee
Therefore, in the limit where $\rmax \to \infty$, the difference of these terms precisely vanishes.

\subsection{Robustness of the CA result}

Here we perform some additional analysis to test the robustness of the results obtained for $\Delta \mathcal{C}_\mathcal{A}$. 

The first possibility we consider here is the addition of a GHY term on the timelike surface $r = r_0$. We emphasize that, strictly speaking, the inclusion of such a term is not warranted. The surface $r = r_0$ is \textit{not} a boundary of the spacetime, as the space degenerates smoothly there and this surface is more akin to an `origin'. The situation is actually somewhat similar to the familiar computation of the Euclidean action in black hole thermodynamics --- a surface term should not included at the Euclidean horizon. Nonetheless, as is clear from the expression for the trace of the extrinsic curvature presented in the previous section, a finite GHY term can arise there and here we simply explore the consequences of this fact. This is because, while the volume form on the $r = r_0 + \epsilon$ surface vanishes as $\epsilon \to 0$, the extrinsic curvature blows up in precisely the right way so that the combination $\sqrt{-h} K$ leaves a finite contribution.  

\begin{figure}[htp]
\centering
\includegraphics[width=0.6\textwidth]{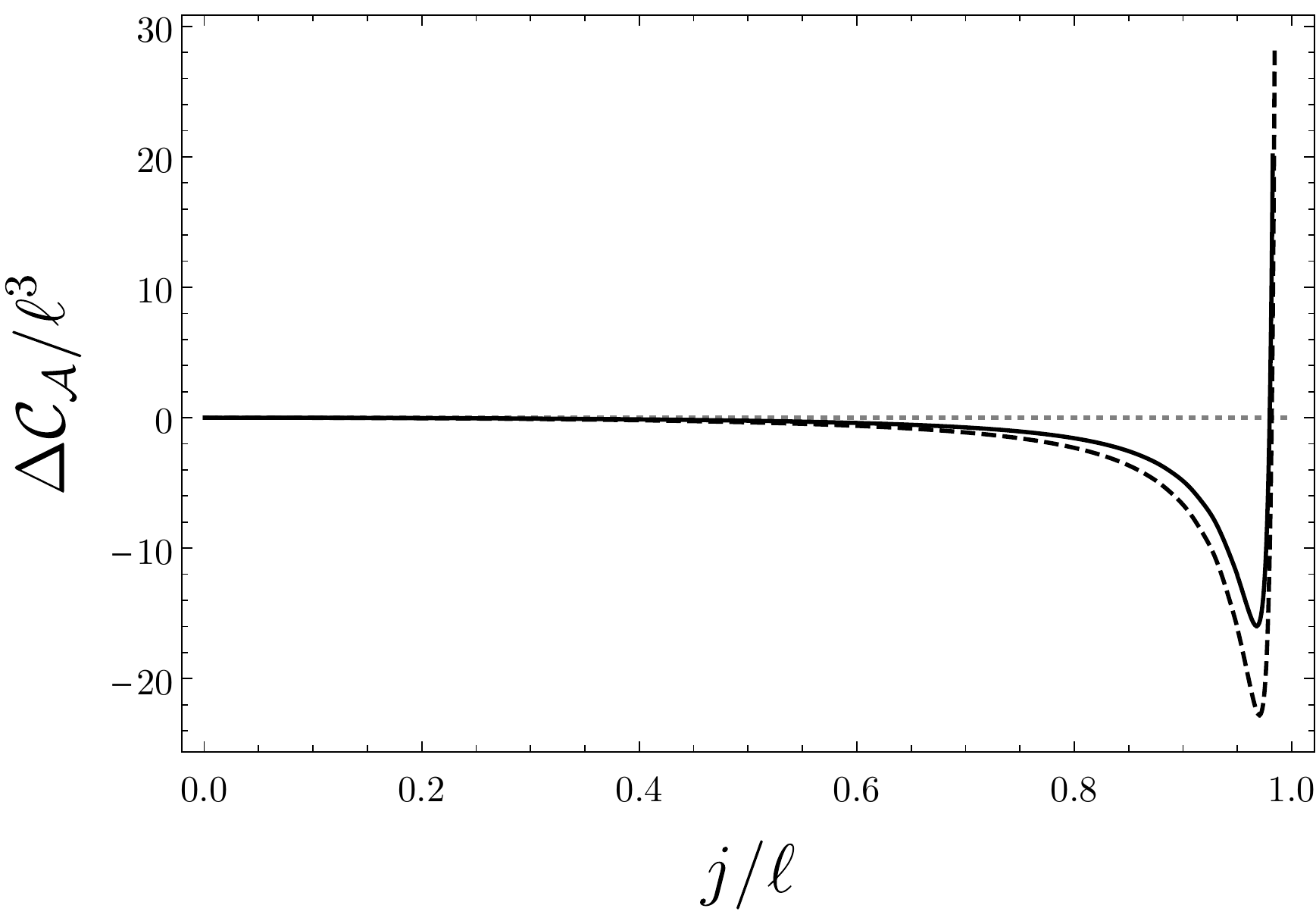}
\caption{A plot of $\Delta \mathcal{C}_\mathcal{A}$ including a GHY contribution at the bubble $r = r_0$. The solid line corresponds to the ``correct'' complexity of formation where this term is neglected, while the dashed curve includes this GHY contribution.}
\label{includingGHY}
\end{figure}

\begin{figure}[htp]
\centering
\includegraphics[width=0.45\textwidth]{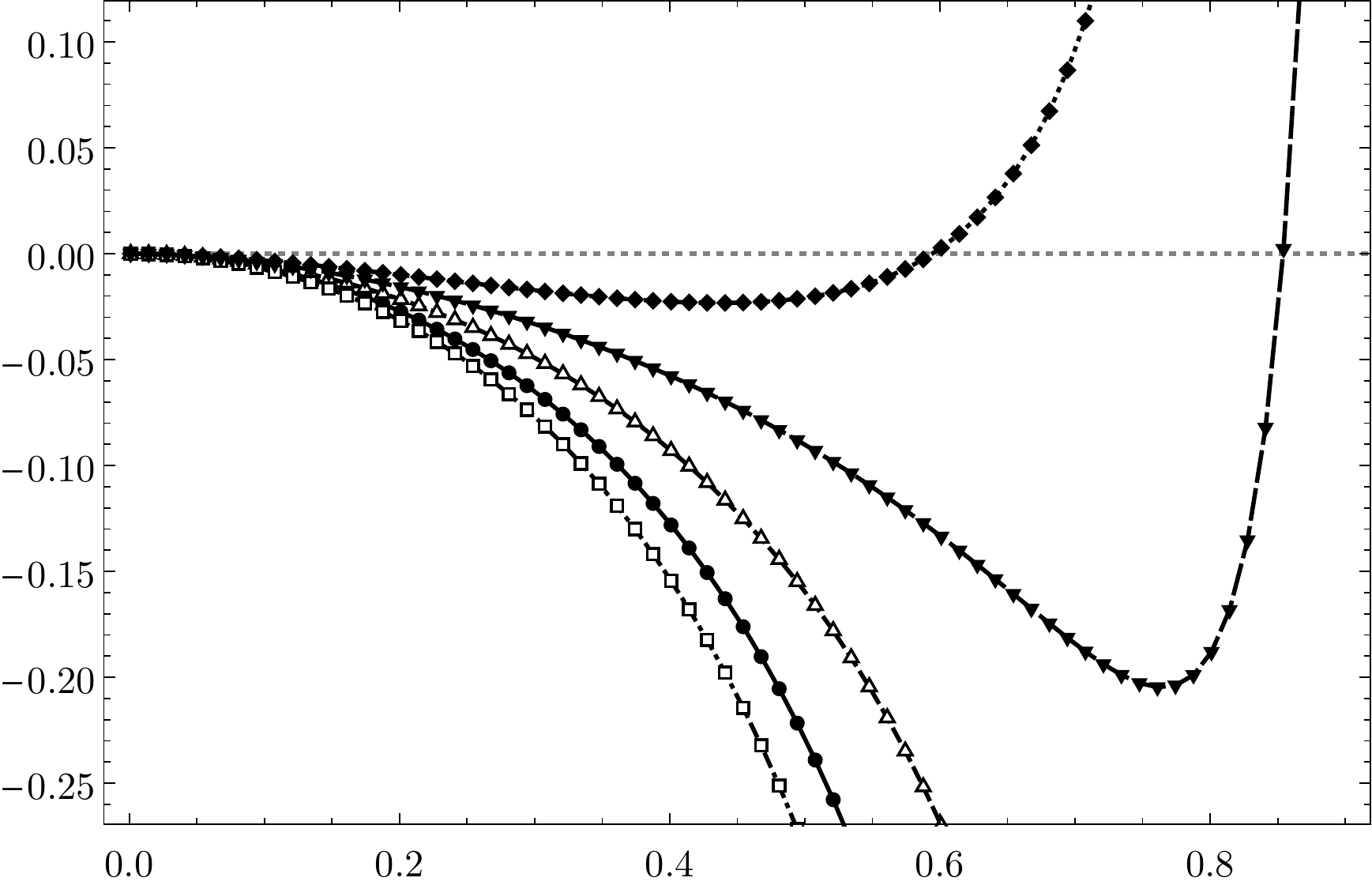}
\quad
\includegraphics[width=0.45\textwidth]{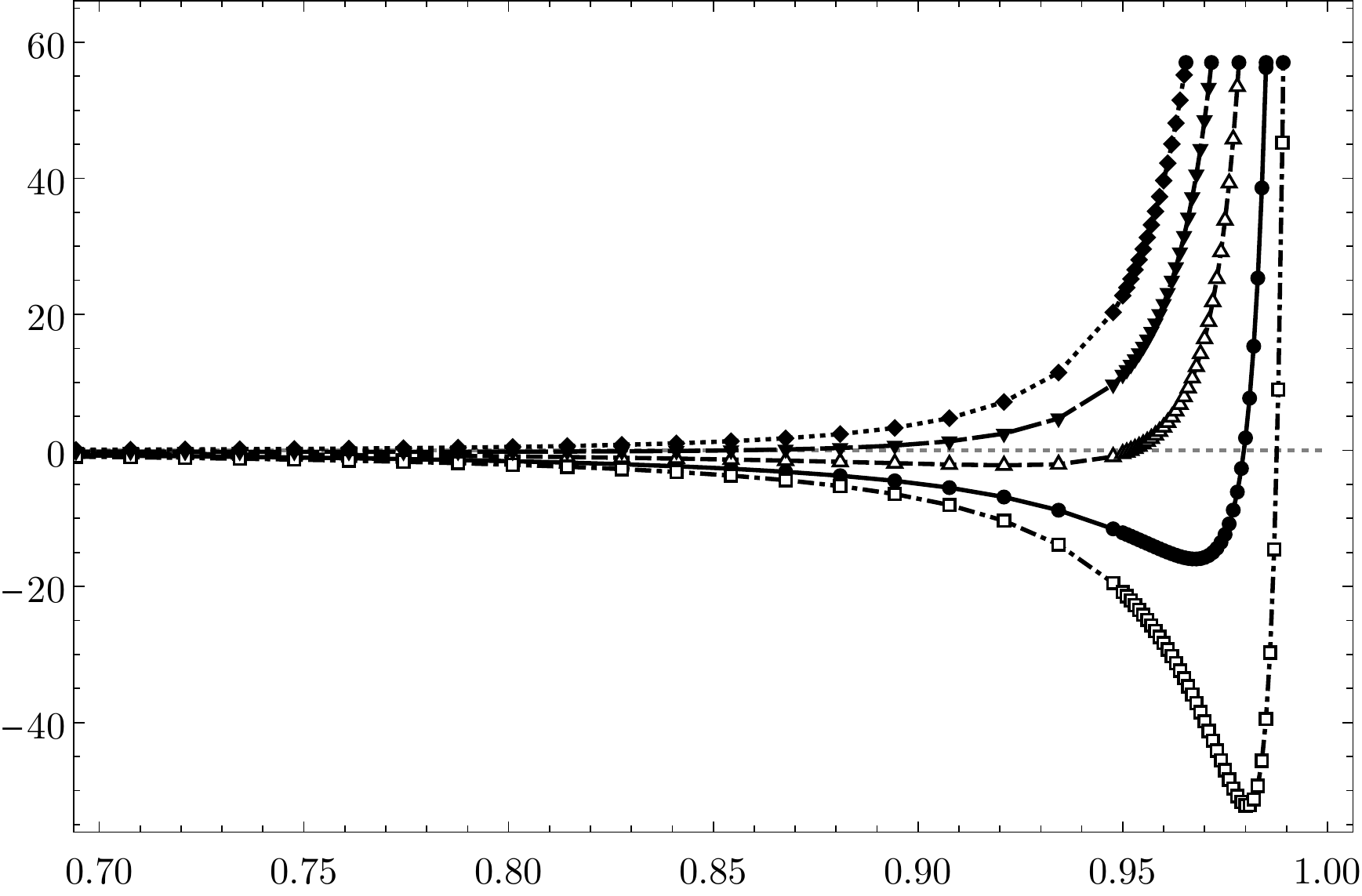}
\caption{Here we show plots of the complexity of formation for different values of $\gamma$. The curves correspond to $\gamma = 3/2, 1, 1/3, -1/3, -1$ in order from bottom to top. Each plot here displays the same five curves, but we have split the plots for clarity.}
\label{gamPlots}
\end{figure}

The GHY term at $r= r_0$ has the finite contribution
\be 
\Delta I_{\rm GHY} = - \frac{\pi r_0^3W'(r_0)}{4 G}  \left[v_\infty - r^*(r_0) \right] \, .
\ee
If this contribution is included in the calculation of the complexity of formation then the result is shown in figure~\ref{includingGHY}.  We see that this term yields a negative contribution to the complexity of formation, lowering the curve. However, it does not result in any qualitative change of behaviour.

The second thing we explore here is the effect of the gauge field on the complexity of formation. To this end, we modify the bulk action in the following way:
\be 
I_{\rm bulk}^\gamma = 
\frac{1}{16\pi G} \int_\mathcal{M}  \star\left (R + \frac{12}{\ell^2}\right)  + \gamma \left[ - 2  F \wedge \star F -\frac{8}{3\sqrt{3}} F \wedge F \wedge A  \right] \, .
\ee
Adjustments such as this to the on-shell action could arise if the boundary conditions on the gauge field were changed from Dirichlet to Neumann or Robin (see, e.g.,~\cite{Goto:2018iay}). The effect of this adjustment is to modify the expression for $\mathcal{I}$ presented in~\eqref{calI} to
\be 
\mathcal{I}(r) \to \mathcal{I}_\gamma(r) = \frac{2 r^4}{\ell^2} - \frac{(3 \gamma -1) r_0^8}{j^2 r^2} \left[\frac{1}{r^2} + \frac{1}{\ell^2} - \frac{1}{j^2} \right] \, .
\ee
We illustrate these effects in figure~\ref{gamPlots}. The electromagnetic terms have the effect of decreasing the complexity of formation for intermediate values of $j$, as evidenced by the fact that positive $\gamma$ pulls the curve down, while negative $\gamma$ shift it up.

\bibliographystyle{JHEP}
\bibliography{Gravities}

\newcommand{\noop}[1]{}

\providecommand{\href}[2]{#2}\begingroup\raggedright\begin{thebibliography}{10}

\bibitem{Lich}
L.~A, \emph{ThŽories de la gravitation et de lՎlectromagnŽtisme relativitŽ
  gŽnŽrale et thŽories unitaires ThŽories Relativistes de la Gravitation et de
  LՎlectromagnŽtisme: RelativitŽ GŽnŽrale et ThŽories Unitaires}.
\newblock Masson et C. Editeurs, Paris, 1955.

\bibitem{Gibbons:1990um}
G.~W. Gibbons, \emph{{Selfgravitating magnetic monopoles, global monopoles and
  black holes}}, \href{http://dx.doi.org/10.1007/3-540-54293-0_24}{\emph{Lect.
  Notes Phys.} {\bfseries 383} (1991) 110--138},
  [\href{https://arxiv.org/abs/1109.3538}{{\ttfamily 1109.3538}}].

\bibitem{Volkov:1998cc}
M.~S. Volkov and D.~V. Gal'tsov, \emph{{Gravitating nonAbelian solitons and
  black holes with Yang-Mills fields}},
  \href{http://dx.doi.org/10.1016/S0370-1573(99)00010-1}{\emph{Phys. Rept.}
  {\bfseries 319} (1999) 1--83},
  [\href{https://arxiv.org/abs/hep-th/9810070}{{\ttfamily hep-th/9810070}}].

\bibitem{Kunduri:2017htl}
H.~K. Kunduri and J.~Lucietti, \emph{{No static bubbling spacetimes in higher
  dimensional EinsteinÐMaxwell theory}},
  \href{http://dx.doi.org/10.1088/1361-6382/aaa744}{\emph{Class. Quant. Grav.}
  {\bfseries 35} (2018) 054003},
  [\href{https://arxiv.org/abs/1712.02668}{{\ttfamily 1712.02668}}].

\bibitem{Shiromizu:2012hb}
T.~Shiromizu, S.~Ohashi and R.~Suzuki, \emph{{A no-go on strictly stationary
  spacetimes in four/higher dimensions}},
  \href{http://dx.doi.org/10.1103/PhysRevD.86.064041}{\emph{Phys. Rev.}
  {\bfseries D86} (2012) 064041},
  [\href{https://arxiv.org/abs/1207.7250}{{\ttfamily 1207.7250}}].

\bibitem{Eperon:2016cdd}
F.~C. Eperon, H.~S. Reall and J.~E. Santos, \emph{{Instability of
  supersymmetric microstate geometries}},
  \href{http://dx.doi.org/10.1007/JHEP10(2016)031}{\emph{JHEP} {\bfseries 10}
  (2016) 031}, [\href{https://arxiv.org/abs/1607.06828}{{\ttfamily
  1607.06828}}].

\bibitem{Keir:2016azt}
J.~Keir, \emph{{Wave propagation on microstate geometries}},
  \href{https://arxiv.org/abs/1609.01733}{{\ttfamily 1609.01733}}.

\bibitem{Bena:2007kg}
I.~Bena and N.~P. Warner, \emph{{Black holes, black rings and their
  microstates}},
  \href{http://dx.doi.org/10.1007/978-3-540-79523-0_1}{\emph{Lect. Notes Phys.}
  {\bfseries 755} (2008) 1--92},
  [\href{https://arxiv.org/abs/hep-th/0701216}{{\ttfamily hep-th/0701216}}].

\bibitem{Kunduri:2014iga}
H.~K. Kunduri and J.~Lucietti, \emph{{Black hole non-uniqueness via spacetime
  topology in five dimensions}},
  \href{http://dx.doi.org/10.1007/JHEP10(2014)082}{\emph{JHEP} {\bfseries 10}
  (2014) 082}, [\href{https://arxiv.org/abs/1407.8002}{{\ttfamily 1407.8002}}].

\bibitem{Horowitz:2017fyg}
G.~T. Horowitz, H.~K. Kunduri and J.~Lucietti, \emph{{Comments on Black Holes
  in Bubbling Spacetimes}},
  \href{http://dx.doi.org/10.1007/JHEP06(2017)048}{\emph{JHEP} {\bfseries 06}
  (2017) 048}, [\href{https://arxiv.org/abs/1704.04071}{{\ttfamily
  1704.04071}}].

\bibitem{Kunduri:2013vka}
H.~K. Kunduri and J.~Lucietti, \emph{{The first law of soliton and black hole
  mechanics in five dimensions}},
  \href{http://dx.doi.org/10.1088/0264-9381/31/3/032001}{\emph{Class. Quant.
  Grav.} {\bfseries 31} (2014) 032001},
  [\href{https://arxiv.org/abs/1310.4810}{{\ttfamily 1310.4810}}].

\bibitem{Qing}
J.~Qing, \emph{{On the uniqueness of AdS spacetime in higher dimensions}},
  {\emph{Annales Henri Poincare.} {\bfseries 5} (2004) }.

\bibitem{Galloway:2002ai}
G.~J. Galloway, S.~Surya and E.~Woolgar, \emph{{On the geometry and mass of
  static, asymptotically AdS space-times, and the uniqueness of the AdS
  soliton}}, \href{http://dx.doi.org/10.1007/s00220-003-0912-7}{\emph{Commun.
  Math. Phys.} {\bfseries 241} (2003) 1--25},
  [\href{https://arxiv.org/abs/hep-th/0204081}{{\ttfamily hep-th/0204081}}].

\bibitem{Herdeiro:2016plq}
C.~A.~R. Herdeiro and E.~Radu, \emph{{Static Einstein-Maxwell black holes with
  no spatial isometries in AdS space}},
  \href{http://dx.doi.org/10.1103/PhysRevLett.117.221102}{\emph{Phys. Rev.
  Lett.} {\bfseries 117} (2016) 221102},
  [\href{https://arxiv.org/abs/1606.02302}{{\ttfamily 1606.02302}}].

\bibitem{Ross:2005yj}
S.~F. Ross, \emph{{Non-supersymmetric asymptotically AdS{$_5 \times S^5$}
  smooth geometries}},
  \href{http://dx.doi.org/10.1088/1126-6708/2006/01/130}{\emph{JHEP} {\bfseries
  01} (2006) 130}, [\href{https://arxiv.org/abs/hep-th/0511090}{{\ttfamily
  hep-th/0511090}}].

\bibitem{Cassani:2015upa}
D.~Cassani, J.~Lorenzen and D.~Martelli, \emph{{Comments on supersymmetric
  solutions of minimal gauged supergravity in five dimensions}},
  \href{http://dx.doi.org/10.1088/0264-9381/33/11/115013}{\emph{Class. Quant.
  Grav.} {\bfseries 33} (2016) 115013},
  [\href{https://arxiv.org/abs/1510.01380}{{\ttfamily 1510.01380}}].

\bibitem{Chong:2005hr}
Z.~W. Chong, M.~Cvetic, H.~Lu and C.~N. Pope, \emph{{General non-extremal
  rotating black holes in minimal five-dimensional gauged supergravity}},
  \href{http://dx.doi.org/10.1103/PhysRevLett.95.161301}{\emph{Phys. Rev.
  Lett.} {\bfseries 95} (2005) 161301},
  [\href{https://arxiv.org/abs/hep-th/0506029}{{\ttfamily hep-th/0506029}}].

\bibitem{Kastor:2010gq}
D.~Kastor, S.~Ray and J.~Traschen, \emph{{Smarr Formula and an Extended First
  Law for Lovelock Gravity}},
  \href{http://dx.doi.org/10.1088/0264-9381/27/23/235014}{\emph{Class. Quant.
  Grav.} {\bfseries 27} (2010) 235014},
  [\href{https://arxiv.org/abs/1005.5053}{{\ttfamily 1005.5053}}].

\bibitem{Creighton:1995au}
J.~D.~E. Creighton and R.~B. Mann, \emph{{Quasilocal thermodynamics of dilaton
  gravity coupled to gauge fields}},
  \href{http://dx.doi.org/10.1103/PhysRevD.52.4569}{\emph{Phys. Rev.}
  {\bfseries D52} (1995) 4569--4587},
  [\href{https://arxiv.org/abs/gr-qc/9505007}{{\ttfamily gr-qc/9505007}}].

\bibitem{Caldarelli:1999xj}
M.~M. Caldarelli, G.~Cognola and D.~Klemm, \emph{{Thermodynamics of
  Kerr-Newman-AdS black holes and conformal field theories}},
  \href{http://dx.doi.org/10.1088/0264-9381/17/2/310}{\emph{Class. Quant.
  Grav.} {\bfseries 17} (2000) 399--420},
  [\href{https://arxiv.org/abs/hep-th/9908022}{{\ttfamily hep-th/9908022}}].

\bibitem{Kunduri:2005zg}
H.~K. Kunduri and J.~Lucietti, \emph{{Notes on non-extremal, charged, rotating
  black holes in minimal D=5 gauged supergravity}},
  \href{http://dx.doi.org/10.1016/j.nuclphysb.2005.07.017}{\emph{Nucl. Phys.}
  {\bfseries B724} (2005) 343--356},
  [\href{https://arxiv.org/abs/hep-th/0504158}{{\ttfamily hep-th/0504158}}].

\bibitem{Kubiznak:2016qmn}
D.~Kubiznak, R.~B. Mann and M.~Teo, \emph{{Black hole chemistry: thermodynamics
  with Lambda}}, \href{http://dx.doi.org/10.1088/1361-6382/aa5c69}{\emph{Class.
  Quant. Grav.} {\bfseries 34} (2017) 063001},
  [\href{https://arxiv.org/abs/1608.06147}{{\ttfamily 1608.06147}}].

\bibitem{Mbarek:2016mep}
S.~Mbarek and R.~B. Mann, \emph{{Thermodynamic Volume of Cosmological
  Solitons}},
  \href{http://dx.doi.org/10.1016/j.physletb.2016.12.042}{\emph{Phys. Lett.}
  {\bfseries B765} (2017) 352--358},
  [\href{https://arxiv.org/abs/1611.01131}{{\ttfamily 1611.01131}}].

\bibitem{Johnson:2014xza}
C.~V. Johnson, \emph{{Thermodynamic Volumes for AdS-Taub-NUT and
  AdS-Taub-Bolt}},
  \href{http://dx.doi.org/10.1088/0264-9381/31/23/235003}{\emph{Class. Quant.
  Grav.} {\bfseries 31} (2014) 235003},
  [\href{https://arxiv.org/abs/1405.5941}{{\ttfamily 1405.5941}}].

\bibitem{Bordo:2019tyh}
A.~B. Bordo, F.~Gray, R.~A. Hennigar and D.~Kubiznak, \emph{{Misner
  Gravitational Charges and Variable String Strengths}},
  \href{http://dx.doi.org/10.1088/1361-6382/ab3d4d}{\emph{Class. Quant. Grav.}
  {\bfseries 36} (2019) 194001},
  [\href{https://arxiv.org/abs/1905.03785}{{\ttfamily 1905.03785}}].

\bibitem{Gunasekaran:2016nep}
S.~Gunasekaran, U.~Hussain and H.~K. Kunduri, \emph{{Soliton mechanics}},
  \href{http://dx.doi.org/10.1103/PhysRevD.94.124029}{\emph{Phys. Rev.}
  {\bfseries D94} (2016) 124029},
  [\href{https://arxiv.org/abs/1609.08500}{{\ttfamily 1609.08500}}].

\bibitem{Ryu:2006bv}
S.~Ryu and T.~Takayanagi, \emph{{Holographic derivation of entanglement entropy
  from AdS/CFT}},
  \href{http://dx.doi.org/10.1103/PhysRevLett.96.181602}{\emph{Phys. Rev.
  Lett.} {\bfseries 96} (2006) 181602},
  [\href{https://arxiv.org/abs/hep-th/0603001}{{\ttfamily hep-th/0603001}}].

\bibitem{VanRaamsdonk:2010pw}
M.~Van~Raamsdonk, \emph{{Building up spacetime with quantum entanglement}},
  \href{http://dx.doi.org/10.1007/s10714-010-1034-0,
  10.1142/S0218271810018529}{\emph{Gen. Rel. Grav.} {\bfseries 42} (2010)
  2323--2329}, [\href{https://arxiv.org/abs/1005.3035}{{\ttfamily 1005.3035}}].

\bibitem{Casini:2011kv}
H.~Casini, M.~Huerta and R.~C. Myers, \emph{{Towards a derivation of
  holographic entanglement entropy}},
  \href{http://dx.doi.org/10.1007/JHEP05(2011)036}{\emph{JHEP} {\bfseries 05}
  (2011) 036}, [\href{https://arxiv.org/abs/1102.0440}{{\ttfamily 1102.0440}}].

\bibitem{Lewkowycz:2013nqa}
A.~Lewkowycz and J.~Maldacena, \emph{{Generalized gravitational entropy}},
  \href{http://dx.doi.org/10.1007/JHEP08(2013)090}{\emph{JHEP} {\bfseries 08}
  (2013) 090}, [\href{https://arxiv.org/abs/1304.4926}{{\ttfamily 1304.4926}}].

\bibitem{Hartman:2013qma}
T.~Hartman and J.~Maldacena, \emph{{Time Evolution of Entanglement Entropy from
  Black Hole Interiors}},
  \href{http://dx.doi.org/10.1007/JHEP05(2013)014}{\emph{JHEP} {\bfseries 05}
  (2013) 014}, [\href{https://arxiv.org/abs/1303.1080}{{\ttfamily 1303.1080}}].

\bibitem{Susskind:2014rva}
L.~Susskind, \emph{{Computational Complexity and Black Hole Horizons}},
  \href{http://dx.doi.org/10.1002/prop.201500093,
  10.1002/prop.201500092}{\emph{Fortsch. Phys.} {\bfseries 64} (2016) 44--48},
  [\href{https://arxiv.org/abs/1403.5695}{{\ttfamily 1403.5695}}].

\bibitem{Susskind:2014moa}
L.~Susskind, \emph{{Entanglement is not enough}},
  \href{http://dx.doi.org/10.1002/prop.201500095}{\emph{Fortsch. Phys.}
  {\bfseries 64} (2016) 49--71},
  [\href{https://arxiv.org/abs/1411.0690}{{\ttfamily 1411.0690}}].

\bibitem{Stanford:2014jda}
D.~Stanford and L.~Susskind, \emph{{Complexity and Shock Wave Geometries}},
  \href{http://dx.doi.org/10.1103/PhysRevD.90.126007}{\emph{Phys. Rev.}
  {\bfseries D90} (2014) 126007},
  [\href{https://arxiv.org/abs/1406.2678}{{\ttfamily 1406.2678}}].

\bibitem{Brown:2015bva}
A.~R. Brown, D.~A. Roberts, L.~Susskind, B.~Swingle and Y.~Zhao,
  \emph{{Holographic Complexity Equals Bulk Action?}},
  \href{http://dx.doi.org/10.1103/PhysRevLett.116.191301}{\emph{Phys. Rev.
  Lett.} {\bfseries 116} (2016) 191301},
  [\href{https://arxiv.org/abs/1509.07876}{{\ttfamily 1509.07876}}].

\bibitem{Brown:2015lvg}
A.~R. Brown, D.~A. Roberts, L.~Susskind, B.~Swingle and Y.~Zhao,
  \emph{{Complexity, action, and black holes}},
  \href{http://dx.doi.org/10.1103/PhysRevD.93.086006}{\emph{Phys. Rev.}
  {\bfseries D93} (2016) 086006},
  [\href{https://arxiv.org/abs/1512.04993}{{\ttfamily 1512.04993}}].

\bibitem{Cai:2016xho}
R.-G. Cai, S.-M. Ruan, S.-J. Wang, R.-Q. Yang and R.-H. Peng, \emph{{Action
  growth for AdS black holes}},
  \href{http://dx.doi.org/10.1007/JHEP09(2016)161}{\emph{JHEP} {\bfseries 09}
  (2016) 161}, [\href{https://arxiv.org/abs/1606.08307}{{\ttfamily
  1606.08307}}].

\bibitem{Huang:2016fks}
H.~Huang, X.-H. Feng and H.~Lu, \emph{{Holographic Complexity and Two
  Identities of Action Growth}},
  \href{http://dx.doi.org/10.1016/j.physletb.2017.04.011}{\emph{Phys. Lett.}
  {\bfseries B769} (2017) 357--361},
  [\href{https://arxiv.org/abs/1611.02321}{{\ttfamily 1611.02321}}].

\bibitem{Yang:2016awy}
R.-Q. Yang, \emph{{Strong energy condition and complexity growth bound in
  holography}}, \href{http://dx.doi.org/10.1103/PhysRevD.95.086017}{\emph{Phys.
  Rev.} {\bfseries D95} (2017) 086017},
  [\href{https://arxiv.org/abs/1610.05090}{{\ttfamily 1610.05090}}].

\bibitem{Cano:2018aqi}
P.~A. Cano, R.~A. Hennigar and H.~Marrochio, \emph{{Complexity Growth Rate in
  Lovelock Gravity}},
  \href{http://dx.doi.org/10.1103/PhysRevLett.121.121602}{\emph{Phys. Rev.
  Lett.} {\bfseries 121} (2018) 121602},
  [\href{https://arxiv.org/abs/1803.02795}{{\ttfamily 1803.02795}}].

\bibitem{Jiang:2018pfk}
J.~Jiang, \emph{{Action growth rate for a higher curvature gravitational
  theory}}, \href{http://dx.doi.org/10.1103/PhysRevD.98.086018}{\emph{Phys.
  Rev.} {\bfseries D98} (2018) 086018},
  [\href{https://arxiv.org/abs/1810.00758}{{\ttfamily 1810.00758}}].

\bibitem{Nally:2019rnw}
R.~Nally, \emph{{Stringy Effects and the Role of the Singularity in Holographic
  Complexity}}, \href{http://dx.doi.org/10.1007/JHEP09(2019)094}{\emph{JHEP}
  {\bfseries 09} (2019) 094},
  [\href{https://arxiv.org/abs/1902.09545}{{\ttfamily 1902.09545}}].

\bibitem{Jefferson:2017sdb}
R.~Jefferson and R.~C. Myers, \emph{{Circuit complexity in quantum field
  theory}}, \href{http://dx.doi.org/10.1007/JHEP10(2017)107}{\emph{JHEP}
  {\bfseries 10} (2017) 107},
  [\href{https://arxiv.org/abs/1707.08570}{{\ttfamily 1707.08570}}].

\bibitem{Chapman:2017rqy}
S.~Chapman, M.~P. Heller, H.~Marrochio and F.~Pastawski, \emph{{Toward a
  Definition of Complexity for Quantum Field Theory States}},
  \href{http://dx.doi.org/10.1103/PhysRevLett.120.121602}{\emph{Phys. Rev.
  Lett.} {\bfseries 120} (2018) 121602},
  [\href{https://arxiv.org/abs/1707.08582}{{\ttfamily 1707.08582}}].

\bibitem{Reynolds:2017jfs}
A.~P. Reynolds and S.~F. Ross, \emph{{Complexity of the AdS Soliton}},
  \href{http://dx.doi.org/10.1088/1361-6382/aab32d}{\emph{Class. Quant. Grav.}
  {\bfseries 35} (2018) 095006},
  [\href{https://arxiv.org/abs/1712.03732}{{\ttfamily 1712.03732}}].

\bibitem{Bombini:2019vuk}
A.~Bombini and G.~Fardelli, \emph{{Holographic entanglement entropy and
  complexity of microstate geometries}},
  \href{https://arxiv.org/abs/1910.01831}{{\ttfamily 1910.01831}}.

\bibitem{Couch:2016exn}
J.~Couch, W.~Fischler and P.~H. Nguyen, \emph{{Noether charge, black hole
  volume, and complexity}},
  \href{http://dx.doi.org/10.1007/JHEP03(2017)119}{\emph{JHEP} {\bfseries 03}
  (2017) 119}, [\href{https://arxiv.org/abs/1610.02038}{{\ttfamily
  1610.02038}}].

\bibitem{Fan:2018wnv}
Z.-Y. Fan and M.~Guo, \emph{{On the Noether charge and the gravity duals of
  quantum complexity}}, \href{http://dx.doi.org/10.1007/JHEP09(2019)121,
  10.1007/JHEP08(2018)031}{\emph{JHEP} {\bfseries 08} (2018) 031},
  [\href{https://arxiv.org/abs/1805.03796}{{\ttfamily 1805.03796}}].

\bibitem{Liu:2019mxz}
H.-S. Liu, H.~Lü, L.~Ma and W.-D. Tan, \emph{{Holographic Complexity Bounds}},
  \href{https://arxiv.org/abs/1910.10723}{{\ttfamily 1910.10723}}.

\bibitem{Sun:2019yps}
W.~Sun and X.-H. Ge, \emph{{Complexity growth rate, grand potential and
  partition function}},  \href{https://arxiv.org/abs/1912.00153}{{\ttfamily
  1912.00153}}.

\bibitem{Chapman:2018dem}
S.~Chapman, H.~Marrochio and R.~C. Myers, \emph{{Holographic complexity in
  Vaidya spacetimes. Part I}},
  \href{http://dx.doi.org/10.1007/JHEP06(2018)046}{\emph{JHEP} {\bfseries 06}
  (2018) 046}, [\href{https://arxiv.org/abs/1804.07410}{{\ttfamily
  1804.07410}}].

\bibitem{Chapman:2018lsv}
S.~Chapman, H.~Marrochio and R.~C. Myers, \emph{{Holographic complexity in
  Vaidya spacetimes. Part II}},
  \href{http://dx.doi.org/10.1007/JHEP06(2018)114}{\emph{JHEP} {\bfseries 06}
  (2018) 114}, [\href{https://arxiv.org/abs/1805.07262}{{\ttfamily
  1805.07262}}].

\bibitem{Agon:2018zso}
C.~A. Agon, M.~Headrick and B.~Swingle, \emph{{Subsystem Complexity and
  Holography}}, \href{http://dx.doi.org/10.1007/JHEP02(2019)145}{\emph{JHEP}
  {\bfseries 02} (2019) 145},
  [\href{https://arxiv.org/abs/1804.01561}{{\ttfamily 1804.01561}}].

\bibitem{Alishahiha:2018lfv}
M.~Alishahiha, K.~Babaei~Velni and M.~R. Mohammadi~Mozaffar, \emph{{Black hole
  subregion action and complexity}},
  \href{http://dx.doi.org/10.1103/PhysRevD.99.126016}{\emph{Phys. Rev.}
  {\bfseries D99} (2019) 126016},
  [\href{https://arxiv.org/abs/1809.06031}{{\ttfamily 1809.06031}}].

\bibitem{Hollands:2007aj}
S.~Hollands and S.~Yazadjiev, \emph{{Uniqueness theorem for 5-dimensional black
  holes with two axial Killing fields}},
  \href{http://dx.doi.org/10.1007/s00220-008-0516-3}{\emph{Commun. Math. Phys.}
  {\bfseries 283} (2008) 749--768},
  [\href{https://arxiv.org/abs/0707.2775}{{\ttfamily 0707.2775}}].

\bibitem{Armas:2011ed}
J.~Armas, P.~Caputa and T.~Harmark, \emph{{Domain Structure of Black Hole
  Space-Times with a Cosmological Constant}},
  \href{http://dx.doi.org/10.1103/PhysRevD.85.084019}{\emph{Phys. Rev.}
  {\bfseries D85} (2012) 084019},
  [\href{https://arxiv.org/abs/1111.1163}{{\ttfamily 1111.1163}}].

\bibitem{Gibbons:2013tqa}
G.~W. Gibbons and N.~P. Warner, \emph{{Global structure of five-dimensional
  fuzzballs}},
  \href{http://dx.doi.org/10.1088/0264-9381/31/2/025016}{\emph{Class. Quant.
  Grav.} {\bfseries 31} (2014) 025016},
  [\href{https://arxiv.org/abs/1305.0957}{{\ttfamily 1305.0957}}].

\bibitem{Cvetic:2005zi}
M.~Cvetic, G.~W. Gibbons, H.~Lu and C.~N. Pope, \emph{{Rotating black holes in
  gauged supergravities: Thermodynamics, supersymmetric limits, topological
  solitons and time machines}},
  \href{https://arxiv.org/abs/hep-th/0504080}{{\ttfamily hep-th/0504080}}.

\bibitem{Compere:2009iy}
G.~Compere, K.~Copsey, S.~de~Buyl and R.~B. Mann, \emph{{Solitons in Five
  Dimensional Minimal Supergravity: Local Charge, Exotic Ergoregions, and
  Violations of the BPS Bound}},
  \href{http://dx.doi.org/10.1088/1126-6708/2009/12/047}{\emph{JHEP} {\bfseries
  12} (2009) 047}, [\href{https://arxiv.org/abs/0909.3289}{{\ttfamily
  0909.3289}}].

\bibitem{Keir:2018hnv}
J.~Keir, \emph{{Evanescent ergosurface instability}},
  \href{https://arxiv.org/abs/1810.03026}{{\ttfamily 1810.03026}}.

\bibitem{Holzegel:2011uu}
G.~Holzegel and J.~Smulevici, \emph{{Decay properties of Klein-Gordon fields on
  Kerr-AdS spacetimes}},
  \href{http://dx.doi.org/10.1002/cpa.21470}{\emph{Commun. Pure Appl. Math.}
  {\bfseries 66} (2013) 1751--1802},
  [\href{https://arxiv.org/abs/1110.6794}{{\ttfamily 1110.6794}}].

\bibitem{Pretorius:1998sf}
F.~Pretorius and W.~Israel, \emph{{Quasispherical light cones of the Kerr
  geometry}}, \href{http://dx.doi.org/10.1088/0264-9381/15/8/012}{\emph{Class.
  Quant. Grav.} {\bfseries 15} (1998) 2289--2301},
  [\href{https://arxiv.org/abs/gr-qc/9803080}{{\ttfamily gr-qc/9803080}}].

\bibitem{Balushi:2019pvr}
A.~A. Balushi and R.~B. Mann, \emph{{Null hypersurfaces in Kerr-(A)dS
  spacetimes}},  \href{https://arxiv.org/abs/1909.06419}{{\ttfamily
  1909.06419}}.

\bibitem{Kastor:2009wy}
D.~Kastor, S.~Ray and J.~Traschen, \emph{{Enthalpy and the Mechanics of AdS
  Black Holes}},
  \href{http://dx.doi.org/10.1088/0264-9381/26/19/195011}{\emph{Class. Quant.
  Grav.} {\bfseries 26} (2009) 195011},
  [\href{https://arxiv.org/abs/0904.2765}{{\ttfamily 0904.2765}}].

\bibitem{Cvetic:2010jb}
M.~Cvetic, G.~W. Gibbons, D.~Kubiznak and C.~N. Pope, \emph{{Black Hole
  Enthalpy and an Entropy Inequality for the Thermodynamic Volume}},
  \href{http://dx.doi.org/10.1103/PhysRevD.84.024037}{\emph{Phys. Rev.}
  {\bfseries D84} (2011) 024037},
  [\href{https://arxiv.org/abs/1012.2888}{{\ttfamily 1012.2888}}].

\bibitem{Chapman:2016hwi}
S.~Chapman, H.~Marrochio and R.~C. Myers, \emph{{Complexity of Formation in
  Holography}}, \href{http://dx.doi.org/10.1007/JHEP01(2017)062}{\emph{JHEP}
  {\bfseries 01} (2017) 062},
  [\href{https://arxiv.org/abs/1610.08063}{{\ttfamily 1610.08063}}].

\bibitem{Gunasekaran:2012dq}
S.~Gunasekaran, R.~B. Mann and D.~Kubiznak, \emph{{Extended phase space
  thermodynamics for charged and rotating black holes and Born-Infeld vacuum
  polarization}}, \href{http://dx.doi.org/10.1007/JHEP11(2012)110}{\emph{JHEP}
  {\bfseries 11} (2012) 110},
  [\href{https://arxiv.org/abs/1208.6251}{{\ttfamily 1208.6251}}].

\bibitem{Lehner:2016vdi}
L.~Lehner, R.~C. Myers, E.~Poisson and R.~D. Sorkin, \emph{{Gravitational
  action with null boundaries}},
  \href{http://dx.doi.org/10.1103/PhysRevD.94.084046}{\emph{Phys. Rev.}
  {\bfseries D94} (2016) 084046},
  [\href{https://arxiv.org/abs/1609.00207}{{\ttfamily 1609.00207}}].

\bibitem{Booth:2001gx}
I.~S. Booth, \emph{{Metric based Hamiltonians, null boundaries, and isolated
  horizons}}, \href{http://dx.doi.org/10.1088/0264-9381/18/20/305}{\emph{Class.
  Quant. Grav.} {\bfseries 18} (2001) 4239--4264},
  [\href{https://arxiv.org/abs/gr-qc/0105009}{{\ttfamily gr-qc/0105009}}].

\bibitem{Parattu:2015gga}
K.~Parattu, S.~Chakraborty, B.~R. Majhi and T.~Padmanabhan, \emph{{A Boundary
  Term for the Gravitational Action with Null Boundaries}},
  \href{http://dx.doi.org/10.1007/s10714-016-2093-7}{\emph{Gen. Rel. Grav.}
  {\bfseries 48} (2016) 94},
  [\href{https://arxiv.org/abs/1501.01053}{{\ttfamily 1501.01053}}].

\bibitem{Cano:2018ckq}
P.~A. Cano, \emph{{Lovelock action with nonsmooth boundaries}},
  \href{http://dx.doi.org/10.1103/PhysRevD.97.104048}{\emph{Phys. Rev.}
  {\bfseries D97} (2018) 104048},
  [\href{https://arxiv.org/abs/1803.00172}{{\ttfamily 1803.00172}}].

\bibitem{Jiang:2018sqj}
J.~Jiang and H.~Zhang, \emph{{Surface term, corner term, and action growth in
  $F(R_{abcd})$ gravity theory}},
  \href{http://dx.doi.org/10.1103/PhysRevD.99.086005}{\emph{Phys. Rev.}
  {\bfseries D99} (2019) 086005},
  [\href{https://arxiv.org/abs/1806.10312}{{\ttfamily 1806.10312}}].

\bibitem{Carmi:2016wjl}
D.~Carmi, R.~C. Myers and P.~Rath, \emph{{Comments on Holographic Complexity}},
  \href{http://dx.doi.org/10.1007/JHEP03(2017)118}{\emph{JHEP} {\bfseries 03}
  (2017) 118}, [\href{https://arxiv.org/abs/1612.00433}{{\ttfamily
  1612.00433}}].

\bibitem{Susskind:2014jwa}
L.~Susskind and Y.~Zhao, \emph{{Switchbacks and the Bridge to Nowhere}},
  \href{https://arxiv.org/abs/1408.2823}{{\ttfamily 1408.2823}}.

\bibitem{Carmi:2017jqz}
D.~Carmi, S.~Chapman, H.~Marrochio, R.~C. Myers and S.~Sugishita, \emph{{On the
  Time Dependence of Holographic Complexity}},
  \href{http://dx.doi.org/10.1007/JHEP11(2017)188}{\emph{JHEP} {\bfseries 11}
  (2017) 188}, [\href{https://arxiv.org/abs/1709.10184}{{\ttfamily
  1709.10184}}].

\bibitem{Horowitz:1998ha}
G.~T. Horowitz and R.~C. Myers, \emph{{The AdS / CFT correspondence and a new
  positive energy conjecture for general relativity}},
  \href{http://dx.doi.org/10.1103/PhysRevD.59.026005}{\emph{Phys. Rev.}
  {\bfseries D59} (1998) 026005},
  [\href{https://arxiv.org/abs/hep-th/9808079}{{\ttfamily hep-th/9808079}}].

\bibitem{Clarkson:2005qx}
R.~Clarkson and R.~B. Mann, \emph{{Eguchi-Hanson solitons in odd dimensions}},
  \href{http://dx.doi.org/10.1088/0264-9381/23/5/005}{\emph{Class. Quant.
  Grav.} {\bfseries 23} (2006) 1507--1524},
  [\href{https://arxiv.org/abs/hep-th/0508200}{{\ttfamily hep-th/0508200}}].

\bibitem{Johnson:2014yja}
C.~V. Johnson, \emph{{Holographic Heat Engines}},
  \href{http://dx.doi.org/10.1088/0264-9381/31/20/205002}{\emph{Class. Quant.
  Grav.} {\bfseries 31} (2014) 205002},
  [\href{https://arxiv.org/abs/1404.5982}{{\ttfamily 1404.5982}}].

\bibitem{Karch:2015rpa}
A.~Karch and B.~Robinson, \emph{{Holographic Black Hole Chemistry}},
  \href{http://dx.doi.org/10.1007/JHEP12(2015)073}{\emph{JHEP} {\bfseries 12}
  (2015) 073}, [\href{https://arxiv.org/abs/1510.02472}{{\ttfamily
  1510.02472}}].

\bibitem{Johnson:2018amj}
C.~V. Johnson and F.~Rosso, \emph{{Holographic Heat Engines, Entanglement
  Entropy, and Renormalization Group Flow}},
  \href{http://dx.doi.org/10.1088/1361-6382/aaf1f1}{\emph{Class. Quant. Grav.}
  {\bfseries 36} (2019) 015019},
  [\href{https://arxiv.org/abs/1806.05170}{{\ttfamily 1806.05170}}].

\bibitem{Brown:2018bms}
A.~R. Brown, H.~Gharibyan, H.~W. Lin, L.~Susskind, L.~Thorlacius and Y.~Zhao,
  \emph{{Complexity of Jackiw-Teitelboim gravity}},
  \href{http://dx.doi.org/10.1103/PhysRevD.99.046016}{\emph{Phys. Rev.}
  {\bfseries D99} (2019) 046016},
  [\href{https://arxiv.org/abs/1810.08741}{{\ttfamily 1810.08741}}].

\bibitem{Goto:2018iay}
K.~Goto, H.~Marrochio, R.~C. Myers, L.~Queimada and B.~Yoshida,
  \emph{{Holographic Complexity Equals Which Action?}},
  \href{http://dx.doi.org/10.1007/JHEP02(2019)160}{\emph{JHEP} {\bfseries 02}
  (2019) 160}, [\href{https://arxiv.org/abs/1901.00014}{{\ttfamily
  1901.00014}}].

\bibitem{Chapman:2018bqj}
S.~Chapman, D.~Ge and G.~Policastro, \emph{{Holographic Complexity for Defects
  Distinguishes Action from Volume}},
  \href{http://dx.doi.org/10.1007/JHEP05(2019)049}{\emph{JHEP} {\bfseries 05}
  (2019) 049}, [\href{https://arxiv.org/abs/1811.12549}{{\ttfamily
  1811.12549}}].

\bibitem{Cooper:2018cmb}
S.~Cooper, M.~Rozali, B.~Swingle, M.~Van~Raamsdonk, C.~Waddell and D.~Wakeham,
  \emph{{Black Hole Microstate Cosmology}},
  \href{http://dx.doi.org/10.1007/JHEP07(2019)065}{\emph{JHEP} {\bfseries 07}
  (2019) 065}, [\href{https://arxiv.org/abs/1810.10601}{{\ttfamily
  1810.10601}}].

\bibitem{Ross:2019rtu}
S.~F. Ross, \emph{{Complexity and typical microstates}},
  \href{http://dx.doi.org/10.1103/PhysRevD.100.066014}{\emph{Phys. Rev.}
  {\bfseries D100} (2019) 066014},
  [\href{https://arxiv.org/abs/1905.06211}{{\ttfamily 1905.06211}}].

\bibitem{Balasubramanian:1999re}
V.~Balasubramanian and P.~Kraus, \emph{{A Stress tensor for Anti-de Sitter
  gravity}}, \href{http://dx.doi.org/10.1007/s002200050764}{\emph{Commun. Math.
  Phys.} {\bfseries 208} (1999) 413--428},
  [\href{https://arxiv.org/abs/hep-th/9902121}{{\ttfamily hep-th/9902121}}].

\bibitem{Brown:1992br}
J.~D. Brown and J.~W. York, Jr., \emph{{Quasilocal energy and conserved charges
  derived from the gravitational action}},
  \href{http://dx.doi.org/10.1103/PhysRevD.47.1407}{\emph{Phys. Rev.}
  {\bfseries D47} (1993) 1407--1419},
  [\href{https://arxiv.org/abs/gr-qc/9209012}{{\ttfamily gr-qc/9209012}}].

\bibitem{Emparan:1999pm}
R.~Emparan, C.~V. Johnson and R.~C. Myers, \emph{{Surface terms as counterterms
  in the AdS / CFT correspondence}},
  \href{http://dx.doi.org/10.1103/PhysRevD.60.104001}{\emph{Phys. Rev.}
  {\bfseries D60} (1999) 104001},
  [\href{https://arxiv.org/abs/hep-th/9903238}{{\ttfamily hep-th/9903238}}].

\end{thebibliography}\endgroup

\end{document}